\documentclass{article}[11pt]
\usepackage{latexsym}
\usepackage[dvips]{graphicx}
\usepackage[english]{babel}
\usepackage{epsfig}
\newcommand \Pomeron {I\!\!P}
\newcommand \jp {$J/\psi \,$}
\newcommand \psip {$\psi' \,$}

\begin{document}
\title{Coherent  Photoproduction from Nuclei} 
\author{
L.~Frankfurt\\
\it School of Physics and Astronomy, Raymond and Beverly Sackler\\
\it Faculty of Exact Science, Tel Aviv University, Ramat Aviv 69978,\\
\it Tel Aviv , Israel\\
M.~Strikman\\
\it Pennsylvania State University, University Park, Pennsylvania 16802\\
M.~Zhalov\\
\it Petersburg Nuclear Physics Institute, Gatchina 188350, Russia}
%\date{}
\maketitle

\begin{abstract}
We argue that study of the cross section of coherent photo(electro)
production of vector mesons off  nuclear targets
provides an effective method to probe  the leading twist 
hard QCD regimes of color transparency and perturbative color opacity 
as well as the onset of black body limit (BBL) in the soft and hard QCD
interactions. In the case of intermediate energies we use
the Generalized Vector Dominance Model  (GVDM) to take into account
coherence effects for two distinctive limits - the soft interactions
for production of $\rho$ and $\rho'$-mesons and  the color transparency
regime for production of charmonium states. We demonstrate that GVDM 
describes very well $\rho$-meson coherent photoproduction at $6 \leq
E_{\gamma} \leq 10$ GeV and predict an oscillating energy dependence
for the  coherent charmonium production.
In the limit of small $x$ we find that hard QCD leads to onset of the
perturbative color opacity even for production of very small onium
states, like $\Upsilon$. The advantages of the process of coherent dijet
photoproduction and hard diffractive processes in general for probing the  
onset of  BBL and measuring the light-cone wave function of the photon
in a hard scattering regime where decomposition over twists becomes 
inapplicable are explained. 
We apply this analysis to the study of the photon induced 
coherent processes in Ultra Peripheral Collisions of ions at LHC and 
demonstrate that the counting rates will be sufficient to study
the physics of color opacity and color transparency at the energies
beyond the reach of the electron-nucleon(nucleus) colliders.

\end{abstract}

\section{INTRODUCTION}

It appears that in the next decade  the photon - nucleus  interactions
will be in the  forefront  of the small {\it x} QCD
dynamics studies. This is due to the   possibilities of the  studying  coherent (and
for some channels incoherent) photon - nucleus interactions 
at energies which exceed at least by a factor of 10 the energies of
electron-nucleon interactions at HERA. Such studies will be feasible
at LHC within the program of the study of the Ultra Peripheral
Collisions (UPC) \cite{felix,baur2002,alice}. This  opens a challenging 
opportunity to get answers to  a number of the fundamental questions 
which could  be investigated in the coherent processes:
how to achieve new QCD regime of strong interaction with small
coupling constant, how interactions 
depend on the type of the projectile and how they change
with an increase of the size/thickness of the target, etc.
Several regimes appear possible depending on 
the incident energy  and the target thickness. A hadronic projectile
(proton, pion, etc)  high-energy interactions with the nucleus  rather rapidly 
approach  the black body limit (BBL) in which the total cross section of the 
interaction is equal to $2\pi R_A^2$ ($R_A \simeq 1.2 \cdot A^{1/3}$). 
Another extreme limit is the interaction of small size projectiles (or wave packages). 
In this case in a wide range of high energies the system remains almost
frozen during the passage through  the nucleus and the regime of 
color transparency is reached in which the interaction of the small
size projectile with a nucleus is rather weak and proportional to A. This 
color transparency phenomenon in hard high energy projectile-nucleus
interactions has been recently observed experimentally in the exclusive 
production of two jets by pions in the coherent diffraction off
nuclei.
At higher energies interactions of small dipole is expected to reach the regime of the
perturbative color opacity. In this regime  the small size
projectile still couples to the gluon field of the target via the 
(skewed) gluon density of the target, $G_{A}(x,Q^2)$,
like in the color transparency regime. However, the scattering
amplitude is not $\propto A$ due to the leading twist (LT)
 shadowing, $G_A/AG_N < 1$. The onset of gluon shadowing  tames somewhat the increase of 
$G_A(x,Q^2)$ between $x\sim 10^{-2}$ and $x\sim 10^{-4}$ and  slows down 
the increase of the small dipole - nucleus interaction with energy.
 The resulting taming is not strong enough to prevent the leading twist
approximation result for the total inelastic cross section from reaching 
and even exceeding the BBL. This violation of unitarity is 
an unambiguous signal that for sufficiently
small $x$ the LT approximation breaks down.

An important practical issue for the studies of the small $x$ dynamics is 
whether the  interactions of small dipoles with several nucleons of
the target are strongly modified  by the leading twist shadowing dynamics \cite{FS99}
or one can neglect the LT gluon shadowing,  like in the model of McLerran and 
Venugopalan\cite{RL}, and focus on the higher twist effects
which are  often modeled in the impact parameter space eikonal model 
\cite{Mueller}. If the leading twist shadowing was  small and
only higher twist effects were taming the  increase of
the dipole - nucleus cross section, the break down of the DGLAP
approximation would occur at rather large $x$. The break down  of
DGLAP may result in  onset of the BBL or taming of the cross sections
at  smaller values. We will argue below that the relative importance of 
leading and higher twist interactions could be experimentally resolved 
using coherent onium photoproduction.

Note also that if the LT gluon shadowing effects were  small enough, 
the BBL for the interaction of $q\bar q$ dipoles of the size $\ge 0.3 - 0.4 fm$ 
could be reached for central collisions with heavy nuclei 
already at ${\it x}\geq 10^{-3}$ that is in the kinematics where
  $\ln {\it x} $ effects in the evolution of the parton densities are small.
  In any case, whatever is the limiting behavior for the interaction of 
the small size dipoles with heavy nuclei, it is of major theoretical interest since it
represents a new regime of interactions  when the leading twist 
approximation and therefore the whole notion of the parton distributions 
becomes inapplicable for the description of hard QCD processes in the small 
${\it x}$ regime. It is worth emphasizing that on the top of providing
higher parton density targets,  nuclei have another important
advantage as compared to the nucleon target. It is a weak dependence of the                               
scattering amplitude on the impact parameter for a wide range                                         
of the impact parameters (in fact, one can combine light and heavy
nuclei to study the dependence of the amplitudes on the nuclear thickness).                                     
On the contrary, in the nucleon case the scattering at large impact
parameters is important even at very small $x$ working to mask the change of the
regime of the interactions at small impact parameters. 

Theoretical studies  of 
 the limiting behavior of
the small dipole - heavy nucleus cross sections did not lead so far to
  definitive results.  It is
conceivable that the QCD dynamics will stop the increase of the cross
section at central impact parameters at the values significantly
smaller than allowed by the BBL. In the following discussion
to emphasize the qualitative difference of the new regime we will use
for simplicity the extreme hypothesis that the impact factor at small
impact parameters reaches the value of one corresponding to the BBL
pattern of the elastic and inelastic cross sections being equal.

There exist many processes where 
the 
projectile wave function is 
 a superposition of configurations of different sizes,  
leading  to the fluctuations of the interaction strength. 
In this respect, interactions of real
and virtual photons with heavy nuclei provide unique opportunities since the 
photon wave function contains both the hadron-like configurations (vector 
meson dominance) and the direct photon configurations (small $q\bar q$ 
components, heavy quark-antiquark components).
The important advantage of the  photon is that at high energies 
the BBL is manifested in diffraction into a multitude of the hadronic final 
states  (elastic diffraction $\gamma \to \gamma $ is negligible)
while in the hadron case only  elastic diffraction survives  in the BBL,
and details of the dynamics responsible for  this regime remain hidden.
Moreover,
 one can {\it post-select} a small size or a large size initial state
 by selecting a particular final state, 
containing for example $c\bar c$  $b\bar b$, or  a leading light meson.
Such
post-selection is much easier in the photon case than in the hadron
case since the  distribution over the sizes of the configurations is much 
broader  in the photon case.

Spectacular manifestations of BBL in (virtual) photon diffraction include
strong enhancement of the large mass tail of the diffractive spectrum as
compared to the expectations of the triple Pomeron limit,  large cross
section of the high $p_t$ dijet production \cite{BBL}.
We emphasize  that the study of the diffractive channels will
allow to distinguish between two extreme scenarios: large suppression
of the cross section due to the leading twist shadowing and nonlinear
regime of the BBL.
Investigation of the coherent diffraction in BBL would allow to perform
unique measurements of various components of the light cone wave function of 
the photon, providing a much more detailed information than similar 
measurements in the  regime where leading twist dominates.

In this review  we  summarize our recent studies of the various 
regimes of the coherent photoproduction off nuclei 
\cite{BBL,FSZpsi,FSZrho,fszstar,fszferrara,fgszch,fgszloch,fgszups}:
the onset of the BBL regime, phenomenon of color transparency and perturbative
color opacity related to the leading twist nuclear gluon shadowing, and the 
pattern of soft QCD phenomena in the proximity to the black body limit.
In addition we consider hard leading twist diffraction off nuclei in DIS.  
We also outline how these effects can be studied in UPC collisions and
provide a comparison with the first UPC data from RHIC.

\section{VECTOR MESON PRODUCTION OFF NUCLEI IN
 THE GENERALIZED VECTOR DOMINANCE MODEL AT INTERMEDIATE ENERGIES}

\subsection{Outline of the model}

Our consideration of the coherent photoproduction processes 
of light flavor and hidden charm 
is based on the use of the eikonal approximation that is the Glauber
model modified to take into account finite longitudinal momentum
transfer
\cite{glauber}. 
The photoproduction cross section $\sigma_{\gamma A\to \rho A}(\omega )$ is given
in the Glauber model by the general expression 
\begin{eqnarray}
\sigma_{\gamma A\to V A}(\omega_{\gamma} )  =\int \limits_{-\infty}^{t_{min}} dt
\frac {\pi} {k_{V}^2}{\left |F_{\gamma A\to V A}(t)\right |}^2=\frac {\pi} {k_{V}^2}
\int \limits_{0}^{\infty} dt_{\bot }
{\left|\frac {ik_{V}} {2\pi}  \int \,d\,{\vec b}\,  e^{i{\vec q}_{\bot }\cdot {\vec b}}
\Gamma ({\vec b}) 
\right|}^2.
\label{crosec}
\end{eqnarray}
Here $\omega_{\gamma} $ is the photon energy, $k_V$is the vector meson momentum,
${\vec q}_{\bot }^{~2}=t_{\bot }={t_{min}-t}$, $-t_{min}=
\frac {M_{V }^4} {4\omega_{\gamma}^2}$ 
is the longitudinal momentum transfer in the $\gamma -V $ transition, 
and $\Gamma ({\vec b})$ is the diffractive nuclear profile function. 
Depending 
on the considered vector meson production process,  the Glauber approach 
can be combined with either the traditional vector dominance model
(VDM) (for the detailed review see Ref. \cite{yenn}) or with   
the generalized vector dominance model (GVDM) \cite{Gribov,Brodsky,GVDM}.
More properly this latter approximation should be called the Gribov-Glauber 
model \cite{Gribovinel} because the space-time evolution of high energy
processes is different in quantum mechanical models and in   quantum field
theory, and therefore theoretical foundations of  the high-energy  model are different.
In particular, in QCD in difference from quantum mechanics a high-energy 
projectile interacts with all nucleons at the same impact parameter 
almost at the same time \cite{Gribovinel}.  
Due to the cancellations between diagonal
and nondiagonal transitions
the GVDM allows to take into
account the QCD  effect of the suppression of interaction of spatially 
small quark-gluon wave packages with a hadron target - the so called
 color transparency phenomenon.
Namely,  cancellations occur between the amplitudes of the photon transition into
a vector state $V_1$ with subsequent conversion to a state $V_2$
and direct production of the state $V_2$, etc. The importance of the
nondiagonal transitions reveals itself in the precocious
Bjorken scaling for moderately small ${\it x}\sim 10^{-2}$ as due to the 
presence in the virtual photon 
of hadron-like  and point-like type configurations  \cite{FS88}.  These 
amplitudes are  also crucial for ensuring a quantitative matching with 
perturbative QCD  regime for $Q^2\leq$ ~few GeV$^2$ \cite{FGS97}. 
The amplitude of the vector meson production off a nucleon
can be written within the GVDM  as
\begin{equation}
A(\gamma + N\to V_j + N )=\sum_{i} {e\over f_{V_i}}A(V_i + N \to V_j + N),
\label{gvdm}
\end{equation}
where $f_{V_i}$ are expressed through $\Gamma(V_{i}\to e^+e^-)$.
Calculation of the vector meson production amplitude off nuclei within the 
Glauber approximation requires taking into  account both  the nondiagonal 
transitions due to the transition of the photon to a different meson
$V'$ in the vertex $\gamma \to V' $ and due to a change of the meson in
multiple rescatterings like $\gamma \to V \to V' \to V$. This physics
is equivalent to inelastic shadowing phenomenon familiar from
hadron-nucleus scattering \cite{Gribovinel}. 
Then  in the optical limit ($A\gg 1$) of the Glauber multistep
production theory one can introduce the eikonal functions ,
$\Phi_{V ,V ^\prime} ({\vec b},z)$, which describe propagation
of the produced objects through the medium and are related to
the diffractive profile function by expression
\begin{eqnarray}
\Gamma ({\vec b})=\lim _{z\rightarrow \infty}\Phi ({\vec b},z).
\label{gamma}
\end{eqnarray}  
Within the optical limit of the Glauber based GVDM 
with accuracy $O(\sqrt \alpha_{em})$ the eikonal functions 
$\Phi_{V ,V ^\prime} ({\vec b},z)$ 
are determined by the solutions of the coupled channel  equations  
\begin{eqnarray}
 \frac {d} {dz} \sum_{V}\Phi _{V }({\vec b},z)=
\sum_{V}  \frac {1} {2ik_V}\biggl [U_{\gamma A\to V A}({\vec b},z)
e^{iq_{\| }^{\gamma \rightarrow V }z}+
\sum_{V^{\prime}} 
U_{V A\to V \prime A}({\vec b},z)
e^{iq_{l}^{V \rightarrow V \prime}z}\Phi _{V ^\prime}({\vec b},z)\biggr ],
\label{cch}
\end{eqnarray}
with the initial condition $\Phi _{V ,V ^\prime} (\vec b ,-\infty)=0$.
The exponential factors $exp [iq_{\| }^{i\rightarrow j}z]$
are responsible for the coherent length effect, $i,j=\gamma ,V ,V ^\prime$, 
 $q_{l }^{i\rightarrow j}=\frac {M_j^2-M_i^2} {2\omega_{\gamma} }$.
The generalized Glauber 
-based optical potentials in the short-range approximation
are given by the expression
\begin{eqnarray}
U_{iA\to jA}({\vec b},z)=-4\pi f_{iN\to jN}(0)\varrho ({\vec b},z).
\end{eqnarray}
Here $f_{iN\to jN}(0)$ are the forward
elementary amplitudes, and $\varrho (\vec b,z) $ 
is the nuclear density normalized by the 
condition $\int d\vec b dz\,\varrho (\vec b,z)=A$.
We calculated $\varrho (\vec b,z)$ in the Hartree-Fock-Skyrme  (HFS) 
model which provided a very good(with an accuracy $\approx 2\%$) description 
of the global nuclear properties of spherical nuclei along the periodical 
table from carbon to uranium\cite{HFS} and the shell momentum distributions 
in the high energy (p,2p)\cite{p2p} and (e,e'p)\cite{eep} reactions. 
If the nondiagonal rescattering amplitudes $f_{VN\to V ^\prime N}=0$,
one can easily integrate  Eq.\ref{cch}, and using the expression for the
elementary amplitude 
$$f_{VN\to VN}(t=0)=\frac {1} {4\pi} k_{V}\sigma_{VN}^{tot}(1-i\alpha_{VN}),$$
obtain the expression for the photoproduction cross section:
\begin{eqnarray}
 {d \sigma_{\gamma A\to VA}\over dt}=
 {d \sigma_{\gamma N\to VN}(t=0)\over dt}
{%\vert 
\left|
\int d^2bdz e^{i{\vec q_t}\cdot {\vec b}}\rho ({\vec b},z)
e^{iq_lz}\cdot e^{-\frac {1} {2} \sigma _{VN}^{tot}({s})(1-i\alpha _{VN})
\int
\limits ^{\infty}_{z}
\rho({\vec b},z')dz'} 
%\vert
\right|
}^2,
\label{dsig}
\end{eqnarray}
well known from early seventies (see for example \cite{yenn}).

\subsection{Production of light vector mesons}
We have used  the 
GVDM
to describe coherent photoproduction of hadronic
states of $M\leq 2$ GeV off nuclei and consider the onset of BBL in
the soft regime\footnote{In our  calculation we neglect the triple
Pomeron contribution which is present at high energies. This
contribution though noticeable for the scattering off the lightest nuclei
becomes a very small correction for the scattering of heavy nuclei due
the strongly absorptive nature of the interaction at the central impact 
parameters.}. In Ref. \cite{Pautz:qm} the simplest nondiagonal model
was considered with two states $\rho$ and $\rho^{\prime}$.
Then the GVDM comprises elementary  amplitudes 
\begin{eqnarray}
f_{\gamma N \to \rho N}=\frac {e} {f_{\rho }} f_{\rho N\to \rho N}+
\frac {e} {f_{\rho ^\prime}}f_{\rho ^\prime N\to \rho N}, 
\nonumber
\\
f_{\gamma N\to \rho ^\prime N}=
\frac {e} {f_{\rho ^\prime}} f_{\rho ^\prime N\to \rho ^\prime N}+
\frac {e} {f_{\rho }}f_{\rho N\to \rho ^\prime N}.
\label{rhogvdm}
\end{eqnarray}

It was assumed that both $\rho$ and $\rho ^\prime$
have the same diagonal amplitudes of scattering off a nucleon.
The ratio of coupling constants was fixed:
$f_{\rho^{\prime}}/f_{\rho}=\sqrt{3}$,
while the ratio of the nondiagonal and diagonal amplitudes
$A(\rho + N \to \rho^{\prime} +N)/A(\rho + N \to \rho +N)=-\epsilon$,
and the value $\sigma^{tot}_{\rho N}$ were found from the fit to
the forward $\gamma +A\to \rho +A$ cross sections
measured at $\omega_{\gamma}=$6.1, 6.6 and 8.8 GeV\cite{MIT}.
One should emphasize here that 
in such GVDM extension $\rho^{\prime}$-meson approximates the hadron production
in the interval of hadron masses $\Delta M^2 \sim 2\, GeV^2$. Thus the values 
of the production cross section refer to the corresponding mass interval.

\begin{figure}
\centering
    \epsfxsize=1.\hsize
     \epsffile{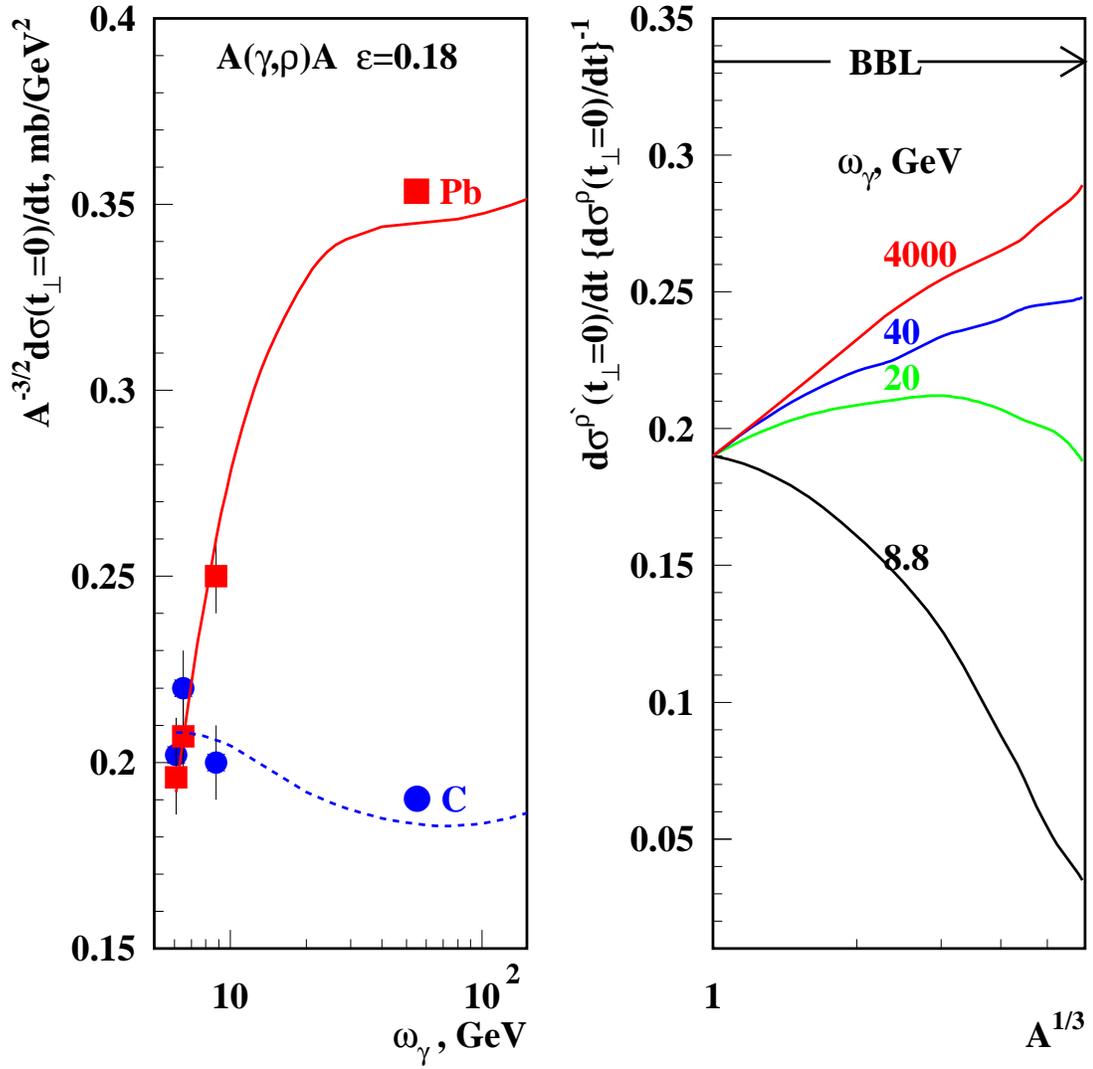}
\caption{The energy dependence of the $\rho$-photoproduction cross section
and the A-dependence of the $\rho^{\prime}/\rho$ photoproduction cross
sections calculated in the GVDM+Glauber model.}
\label{rhoen}
\end{figure}

We refined this model in \cite{FSZrho}. The dependence on the nuclear 
structure parameters was diminished by calculating the nuclear densities
in the Hartree-Fock-Skyrme  (HFS) approach. Next, we  used in all our
calculations the parameterization of ~\cite{LD} for the  $\rho N$ amplitude 
which was obtained from the fit to the experimental data on photoproduction
off the proton target. The value of  $\epsilon $ was fixed at 0.18 to ensure 
the best fit
%(Fig.\ref{figrhoan}) 
of the measured differential cross section of
the $\rho$-meson photoproduction off lead at $\omega_{\gamma}=6.2$ GeV and
$t_{\bot }=0.001$ $GeV^2$.  Note that this value of $\epsilon$ leads to a 
suppression of the differential cross section of the $\rho$-photoproduction 
in $\gamma +p\to \rho +p$ by a factor of $(1-\epsilon/\sqrt{3})^2\approx 0.80$
practically coinciding with phenomenological  renormalization factor 
${ R=0.84}$ introduced in \cite{LD} to achieve the best fit of the elementary 
$\rho$-meson photoproduction forward cross section in the VDM which neglects 
mixing effects.
With all parameters fixed we calculated the  differential cross sections 
of $\rho$-production off nuclei and found a good agreement with 
the data \cite{MIT}, see a detailed comparison in \cite{FSZrho}.
In view of a good agreement of the model with the data
on $\rho$-meson production in the low energy domain we  used this model
to consider the $\rho$-meson photoproduction at higher energies of
photons. The increase of the coherence length  with the photon energy
leads to a  qualitative difference in the energy dependence of
the coherent vector meson production off light
and heavy nuclei  and to a  change of the A-dependence
for the ratio of the forward $\rho^{\prime}$ and $\rho$-meson production
 cross sections between $\omega_{\gamma} \sim 10$ GeV 
and $\omega_{\gamma}\sim 40$ GeV (Fig.\ref{rhoen}).  
The observed pattern reflects the difference of  the 
coherence lengths of the $\rho$-meson and a heavier $\rho^\prime$-meson
which is important for the intermediate photon energies $\leq 30$ GeV.
The corrections due to nondiagonal transitions are relatively  small 
($\sim$ 15\%) for the case of  $\rho$ production off a nuclei. As a
result, we find that the GVDM cross section is close to the one calculated 
in the VDM for heavy nuclei as well. Situation for $\rho^{\prime}$ production
is much more interesting. 
  The cross section of $\rho^{\prime}$ 
production off a nucleon is strongly suppressed as compared to the  case 
when the $\rho \leftrightarrow  \rho^{\prime}$  transitions are switched off.
The extra suppression factor is $\approx 0.5$.
The non-diagonal
transitions disappear in the limit of large A 
(corresponding to the BBL)  due to 
the condition of orthogonality of hadronic wave functions in
 accordance with the general argument of Gribov 
\cite{Gribov}. 
Therefore in the limit of $A\to \infty$ we expect that the relation
\begin{equation}
{d\sigma (\gamma + A \to h_1 +A )/dt \over
d\sigma (\gamma + A \to h_2 +A )/dt}_{\left|A\to \infty\right.}
={\sigma(e^+e^-\to h_1)\over \sigma(e^+e^-\to h_2)}
\approx \left(f_2/f_1\right)^2,
\label{bbl1}
\end{equation}
should be fulfilled for the productions of states $h_1, h_2$ of invariant 
masses $M_1^2,M_2^2$ at $t_{\bot }=0$.
Indeed we have found from calculations that in the case of the coherent 
photoproduction off lead the nondiagonal transitions becomes strongly
suppressed with increase of the photon energy. As  a result  the 
$\rho^{\prime}/\rho$ ratio increases, exceeds the ratio of the 
$\gamma p\to Vp$ forward cross sections calculated with accounting for 
$\rho\leftrightarrow \rho^{\prime}$ transitions already at $\omega_{\gamma}\geq 50$ GeV 
and becomes close to  the value of $f^2_{\rho}/f^2_{\rho^{\prime}}$ which 
can be considered as the limit when one can treat the interaction with
the heavy nucleus as a  black one. 
It is worth noting here that presence of nondiagonal transitions
 in terms of the formalism of the scattering eigen states
\cite{GW} 
corresponds to the fluctuations of the values of the
interaction cross sections  for the real photon.
The GVDM  discussed in the paper leads to small ($\propto 10\%$) color 
transparency effects at intermediate energies for the cross section of
semiinclusive photoproduction processes. Really,  this model corresponds
to the propagation of states with cross sections:
$\approx \sigma(VN)(1 \pm \epsilon)$ . In the case of electroproduction 
$\epsilon$ should be significantly larger: 
$$\epsilon\approx{f_{\rho'}\over f_{\rho}}=
\sqrt{{\Gamma(\rho\rightarrow e^+ e^-)\over m_{\rho}}/
{\Gamma(\rho'\rightarrow e^+ e^-)\over m_{\rho'}}}.$$
This equation follows from Eq.(\ref{gvdm}) where 
the left hand side
is put to zero because the cross section of 
the elastic vector meson electroproduction 
rapidly decreases with  $Q^2$. The presence of the CT phenomenon within 
the GVDM leads to a substantial modification of the  pattern of the approach 
to BBL. 
The nondiagonal transitions become more 
important
with the  increase of $Q^2$,  
 leading to an enhancement of the effects discussed above. 
In particular, the fluctuations of strengths of
interaction would lead at large $Q^2$ to the color transparency
phenomenon. Presence in GVDM of significantly different  masses  of
$\rho, \rho', ...$ makes it
impossible to describe all fluctuations of strengths
of interaction in terms of one coherent length.

\subsection{Photoproduction of $J/\psi,\psi'$-mesons  at intermediate energies.}

There is a qualitative difference between GVDM description of the
light meson production described above 
and production of charmonium states. 
In the case
of the photoproduction of light mesons the soft physics dominates 
both at low and high energies.
Hence, the  Gribov-Glauber model should be applicable in a wide
range of energies. At the same time the space-time picture of the
process
changes with energy. At low energies
the meson is formed at the distances smaller than
 the typical interaction length of a meson in the nucleus.  
With the increase of energy the formation length starts to exceed the
nuclear size and 
a photon converts 
to a system of $q\bar q$ and higher Fock-components before the
target,  and one has to account for the interactions of this system with the
media and subsequent transition of the system to a vector meson.

The situation is much more involved in the case of (charm)onium production.
At high energies both the coherence length 
$l_c\approx {{2\omega_{\gamma}}m_V^{-2}}$ and the formation length 
$l_f \approx 2\omega_{\gamma} {[m_{\psi'}^2-m_{J/\psi}^2]}^{-1}$ 
(distance on which the squeezed $q\bar q$ pair transforms into the 
ordinary meson) are large and the color transparency
phenomenon reveals itself. In particular, it  explains
 the fast increase of the cross section
with energy observed at HERA (for  reviews of small x phenomena 
 see Ref.~\cite{Badelek:gs,hera}). Consequently,
the value of the cross section extracted from the charmonium photoproduction
characterizes the interaction of the squeezed $c\bar c$ pair with a 
nucleon rather than the charmonium-nucleon interaction.
In the high energy limit the very  small interquark distances in the 
wave function of the photon dominate, and one has to treat
the interaction of small dipoles with nuclei. In this case the eikonal
approximation gives a qualitatively wrong answer since it does not take into
account the leading twist effect of the gluon shadowing(see the detailed
analysis in~\cite{FGMS2002}), while the leading twist analysis predicts large
shadowing effects~\cite{FSZpsi}, see discussion in section 3.
 On the other hand, at the intermediate
energies when onium states are formed inside the
nucleus the nonperturbative effects at a transverse distance scale
comparable to the charmonium size becomes important.  
The hadronic basis description would be more relevant 
in this case. However the VDM which takes into account 
only diagonal vector meson transitions does not account properly
for
the basic QCD dynamics of interaction.  In particular, SLAC
data~\cite{slac2} show that $\sigma_{\gamma+N\rightarrow \psi'N}\approx 
0.15 \cdot \sigma_{\gamma+N\rightarrow \psi N}
$ which within VDM corresponds
to:  $\sigma_{\psi' N} /\sigma_{J/\psi N}\approx 0.7$. This conclusion is in
evident contradiction with the QCD expectation that the hadron
interaction cross section 
should be scaled approximately as the transverse area occupied by color:
$\sigma_{\psi' N}/\sigma_{J/\psi N} \propto r_{\psi'}^2 /r_{J/\psi}^2$. 
Thus in  this naive QCD picture 
$\sigma(\psi' N)/\sigma(J/\psi N) \sim 4$.
 A QCD explanation of such
failure of VDM is based on the observation \cite{FS91,kophuf} that in 
photoproduction of both the \jp and \psip mesons the small relative distances
$\sim 1/m_c$ dominate in the $c\bar c$ component of the photon wave
function. Therefore the  cross sections which enter into the ratio of \jp and
\psip yields are cross sections of the interaction of  the small
dipoles,
not genuine  mesons. 
The suppression of the production of \psip in this picture is
primarily due to a smaller leptonic decay width of \psip (a factor of
1/3). A significant  additional suppression  comes from importance  of more massive
$c\bar c$ intermediate states in the photon wave function 
($\ge M_{J/\psi}, M_{\psi'}$ respectively).
The exact value of the suppression is sensitive to the 
details of the onium wave functions and the dependence 
of the dipole - nucleon cross section on the size of the dipole
\cite{FKS,Suzuki:2000az}.
%%%

The dominance of small
$c\bar c$ configurations in photoproduction processes is relevant for
the significant probability of nondiagonal 
$J/\psi \leftrightarrow  \psi ^\prime$ diffractive transitions. 
The GVDM adjusted 
 to account for 
the color screening phenomenon~\cite{FS91,kophuf}
allows to take into account 
QCD dynamics using a hadronic basis. Such a description is limited to
 the regime 
of small coherence lengths (medium energies), 
where leading twist shadowing is not important.
We use the GVDM to consider the coherent photoproduction of hidden
charm mesons off nuclei at moderate photon energies $20\mbox{ GeV} \leq
\omega_{\gamma} \leq 60\mbox{ GeV}$ where the coherence length for the
$\gamma V$ transition $l_{c}$ is
still close enough to the internucleon distance in nuclei while the
formation length $l_{f}$ is comparable to the radii of heavy nuclei. In this
energy range produced charmonium states have a noticeable probability
to rescatter in sufficiently heavy nuclei.
Therefore,
one would be able to reveal the fluctuation of the charmonium-nucleon
interaction strength as due to the diagonal $\psi N\rightarrow \psi N$ and
nondiagonal $\psi N\Leftrightarrow \psi' N$ rescatterings for moderate
energies. 
The GVDM which we outlined  above 
takes into account the coherence
length effects via the Glauber model approximation. 
The key distinction from the $\rho-$meson case
is choosing  the parameters of the model 
 to account for the space-time evolution of
spatially small $c\bar c$ pair.
It is
important that the inelastic shadowing corrections
related to the production of higher mass states~\cite{Gribovinel} are
still insignificant. 
A reasonable 
starting approximation to evaluate the amplitude of the charmonium-nucleon 
interaction is to restrict ourselves to the basis of \jp and 
\psip states for the photon wave function in Eq.\ref{gvdm}.
Then, similarly to the considered above case of $\rho, \rho'$ production
we have two equations comprising six elementary amplitudes.
The charmonium-nucleon coupling constants: 
${{f_{J/\psi}^2}\over {4\pi}}=10.5\pm 0.7$, 
and ${{f_{\psi ^{\prime}}^2}\over {4\pi}}=30.9\pm 2.6$ are 
determined from the widths 
of the vector meson decays $V \rightarrow e\bar e$.
Since in the photoproduction
processes $c\bar c$ pair is produced within the spatially small configuration
one can neglect for a moment by the direct photoproduction amplitude and
obtain the approximative relations between rescattering amplitudes

\begin{eqnarray}
f_{\psi^{\prime} N\to \psi ^{\prime} N}\approx -{f_{\psi^{\prime}}\over f_{J/\psi}}
f_{\psi^{\prime} N\to J/\psi N}
\approx {f_{\psi^{\prime}}^2\over f_{J/\psi}^2}f_{J/\psi N\to J/\psi N} .
\end{eqnarray}
Large values for nondiagonal amplitudes 
$f_{\psi^{\prime} N\to J/\psi N}\approx -1.7 f_{J/\psi N\to J/\psi N}$
are a characteristic QCD property of hidden charm and
beauty meson-nucleon interaction. Note that the negative sign of the
nondiagonal amplitude is dictated by the QCD factorization theorem. A
positive sign of the forward photoproduction $f_{\gamma N\rightarrow
J/\psi(\psi') N}$ amplitudes as well as the signs of the coupling constants
$f_{J/\psi}$ and $f_{\psi'}$ are determined by the signs of the charmonium
wave functions at r=0. From the approximative estimates above it also 
follows that $\sigma_{\psi ^{\prime} N}\approx 9\sigma_{J/\psi N}$. 
This is much larger  than  $\sigma_{\psi' N}\sim$ 20 mb~ suggested by  analyses
of  the data on $\psi'$ absorption 
in nucleus-nucleus collisions
\cite{vmvogt,kharzeev}. When combined with  the SLAC
data~\cite{slac1}, this  corresponds to 
$\sigma_{\psi' N}/\sigma_{J/\psi N}\approx
5\div 6$ with large experimental and theoretical errors.

To fix elementary amplitudes more accurately within GVDM (see %for details
\cite{fgszch} for details) we parameterized the elementary photoproduction cross
section in the form used by the experimentalists of HERA to describe their
data. This form has no firm theoretical justification but it is convenient
for the fit 
\begin{equation} 
\sigma_{\gamma+N \rightarrow V+N} \propto F_{2g}^2(t) \left({s\over 
s_0}\right)^{0.4}\quad. 
\end{equation} 
Here $s=2\omega_{\gamma}m_N$ is the invariant energy for the photon scattering 
off a free nucleon, 
$s_0=40\, GeV^2$ is the reference point and $F_{2g}(t)$
is the two-gluon form factor of a nucleon.

We also used the SLAC results for the 
forward $\gamma N \to J/\psi N$ cross section~ \cite{slac2}
and for the ratio
\begin{equation}
\left.{d\sigma_{\gamma N\rightarrow \psi \prime N}\over dt}\right|_{t=t_{min}}=
0.15\cdot\left.{d\sigma_{\gamma N\rightarrow J/\psi N}\over dt}\right|_{t=t_{min}}.
\end{equation}
Besides, we fixed the $J/\psi N$ cross section 
$\sigma_{J/\psi N}\approx (3.5\pm 0.8)$ mb as measured at
SLAC\cite{slac1} and used the reasonable assumption about the energy dependence
of this cross section as the sum of soft and 
hard physics:
\begin{equation}
\sigma_{J/\psi N}=3.2\mbox{ mb}\left({s\over s_{0}}\right)^{0.08} +
0.3\mbox{ mb} 
\left({s\over s_{0}}\right)^{0.2}\quad.
\label{psiN}
\end{equation}
The existence of a hard part of the $J/\psi$N cross section is consistent
with the GVDM, because the photoproduction amplitude has a stronger energy
dependence than the Pomeron exchange, i.e.\ soft scattering amplitudes. 
 Next, we used  this input, the optical theorem and the well known
Gribov-Migdal relation
\begin{equation}   
{\Re f_{\Psi N\to  \Psi N}}   
={s\pi\over 2}{{\partial \over \partial    
\ln{s}}{\Im f_{\Psi N\to \Psi N}\over s}}\quad,
\label{GM}  
\end{equation}   
to  determine from  Eq.\ref{gvdm} all elementary amplitudes in the 
discussed energy range.
In particular, we found the value
$\sigma_{\psi' N}\approx 8 $ mb. 
However, the input parameters of the 
GVDM, namely 
the experimental cross sections of the forward elementary 
photoproduction and, especially, the value of $\sigma_{J/\psi N}$,
 are known with large uncertainties.
In result, we obtained  elementary cross sections changing in the
ranges: $2.5\,mb\leq  \sigma_{J/\psi N}\leq  5\,mb$ and
$6\,mb\leq  \sigma_{\psi ^\prime N}\leq  12\,mb$, in the discussed
energy range. We checked
how a variation of $\sigma_{J/\psi N}$ within the experimental errors
influences our results. 

\begin{figure}
    \centering
        \leavevmode
        \epsfxsize=.9\hsize
        \epsffile{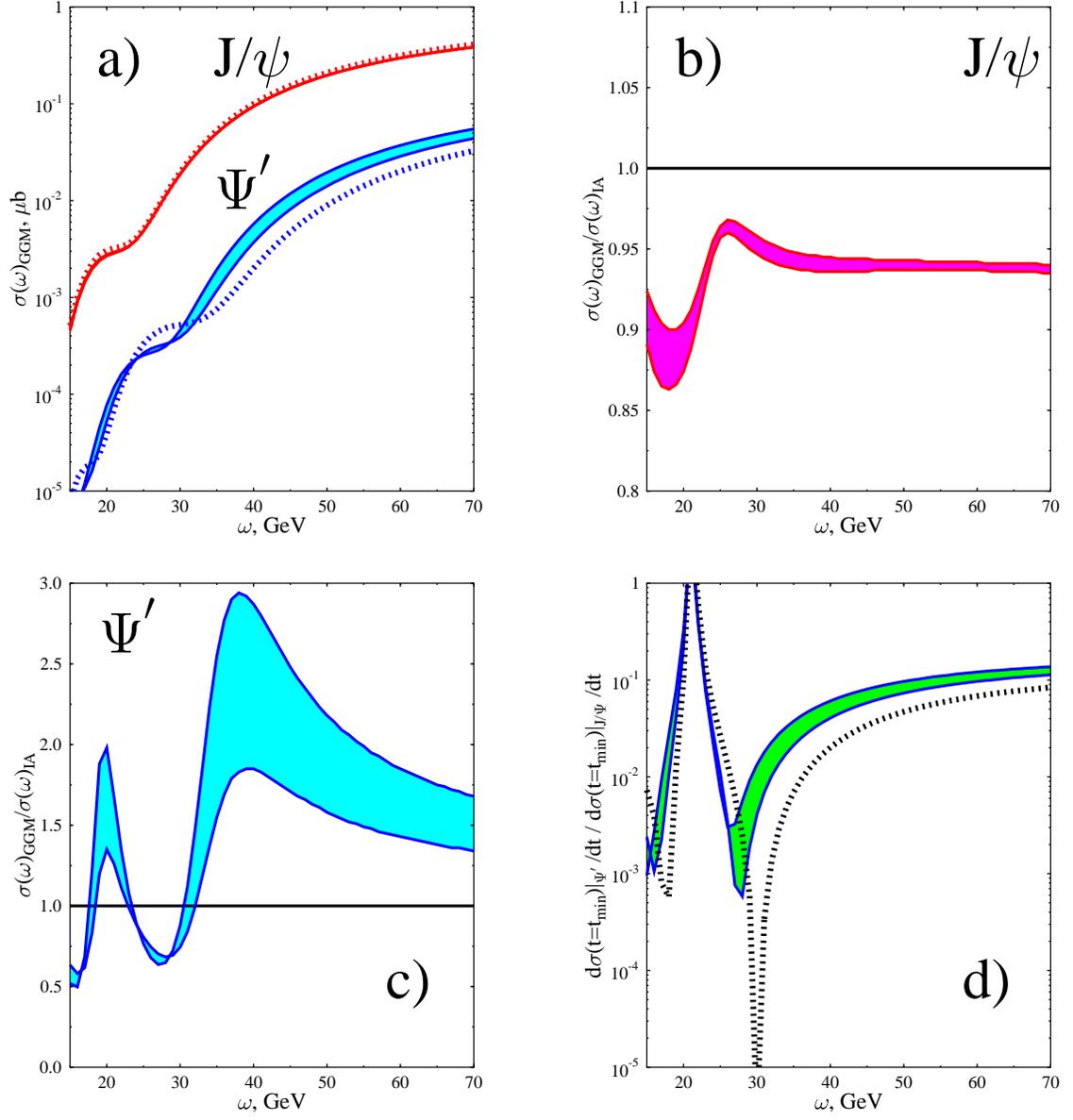}
\caption{The coherent photoproduction of charmonia off Ca:
a). Energy dependence of photoproduction cross sections;
b). Ratio of cross sections calculated within 
the GVDM and IA for $J/\psi$;
c). The same for $\psi ^\prime$;
d). Ration of the forward $\psi ^\prime$-production cross section to that
for $J/\psi$.  
The filled areas show the variation
of the cross sections due to the uncertainty of the experimental $J/\psi$N 
cross section.}
\label{jpsipsi}
\end{figure}

\begin{figure}
    \centering
        \leavevmode
        \epsfxsize=1.\hsize
       \epsffile{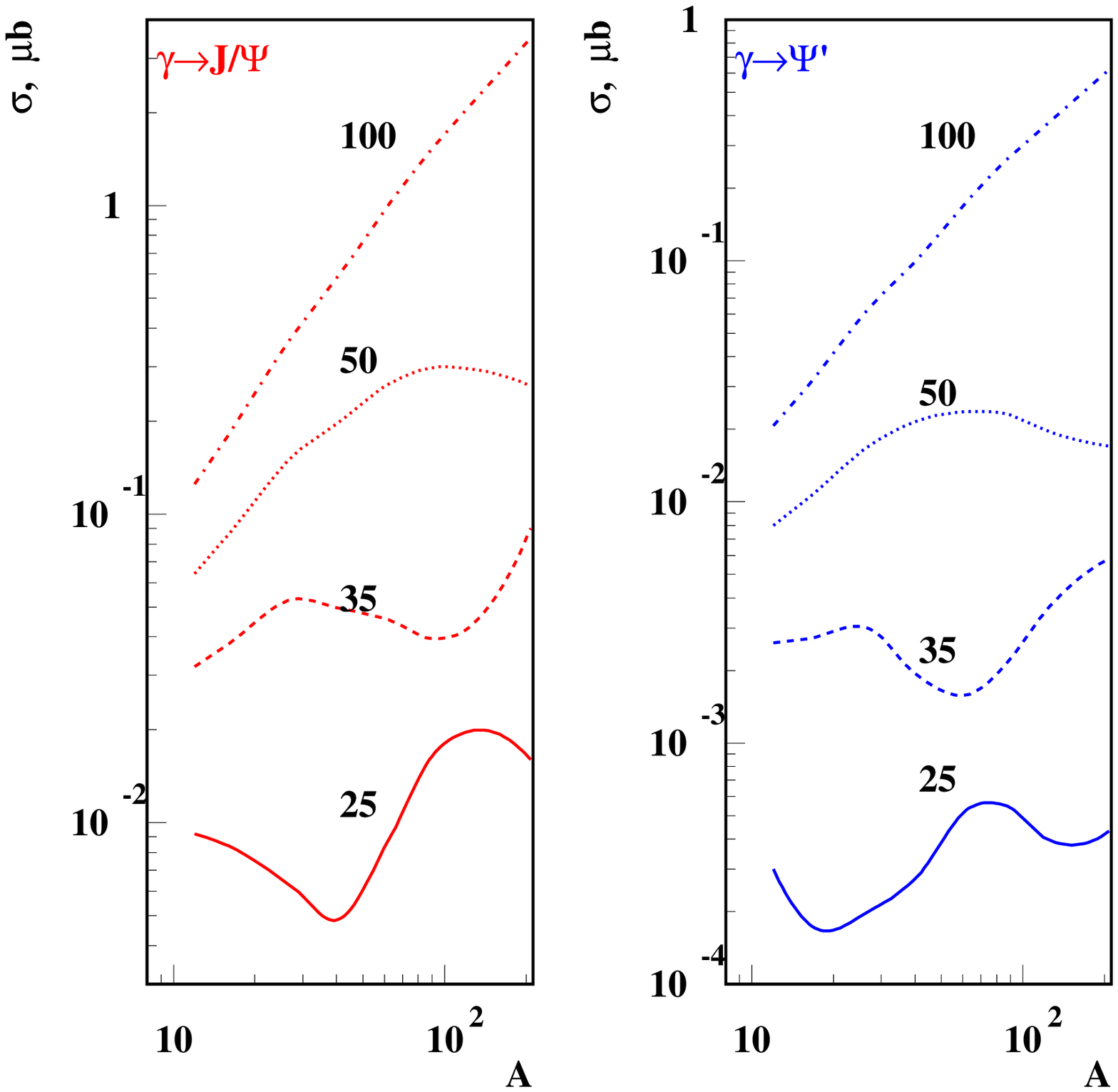}
\caption{
The A-dependence of the  coherent charmonium 
photoproduction cross sections calculated in the Generalized Vector Dominance 
Model for different energies of the photon.
}
\label{adpsi}
\end{figure}

In our  papers\cite{fgszch},\cite{fgszloch} we analyzed the coherent 
photoproduction of
charmonia off light(Si) and heavy(Pb) nuclei.        
Here we present the cross
sections of the coherent photoproduction of $J/\psi$ and $\psi'$ off Ca.  
The energy dependence of these cross sections is compared 
(Fig.\ref{jpsipsi}a) 
to that obtained in the Impulse Approximation where all rescatterings 
of the produced vector mesons are neglected, and the cross section is given
by the simple formula
\begin{eqnarray}
{d\sigma_{\gamma A\to VA}(s,t )\over {d\,t}}=
{d\sigma_{\gamma N\to VN}(s,t_{min})\over dt}\cdot
{\left |
{\int \limits_{0}^{\infty}e^{i{\vec q}_{\bot }\cdot \,{\vec b}}
d\,{\vec b}\int \limits_{-\infty}^{\infty}d\,z\,
e^{iz\cdot q_{\| }^{\gamma V }}\varrho  ({\vec b},z)}
\right |}^2.
\label{iacs}
\end{eqnarray}

The distinctive feature of the coherent charmonium
photoproduction is that differential cross sections  oscillate as a function
of  the photon
energy. The major source for such a behavior at intermediate
energies is oscillating behavior of the 
longitudinal nuclear form factor at the relatively large values
 of $t_{min}$ in 
the photoproduction vertex. 

The cross section of the \jp photoproduction in the  GVDM is close to
that calculated in the Impulse Approximation - the shapes of
curves are very similar and the values of the cross sections are only slightly
reduced at energies below 40 GeV(Fig.\ref{jpsipsi}b). This indicates
that it would be very difficult to extract the total $J/\psi N$ cross
section from such a measurement.
At the same time, since the 
the nuclear form factor is known with a high precision 
from the high energy elastic
electron-nucleus scattering, one can use the  
coherent \jp
photoproduction off the spherical nuclei  to
determine the elementary
$\gamma N\to J/\psi N$ amplitude in a wide range of energies for $t \sim 0$.
 
The picture is qualitatively different for the coherent \psip
photoproduction off nuclei. Due to the higher threshold
($E_{th}^{\psi'}-E_{th}^{J/\psi})\approx 3$ GeV,
 the direct \psip production
off nuclei is suppressed 
in the impulse approximation (for the same incident  energy)
as compared to the  \jp production by
the nuclear form factor.  Since $q_{\| }^{\gamma
\psi'}\approx {{m_{\psi'}^2}\over {m_{J/\psi}^2}} q_{\| }^{\gamma J/\psi} $,
the minima of the impulse approximation distribution are shifted. The
contribution of the nondiagonal term $\gamma\to J/\psi\to \psi'$ essentially
increases the \psip yield  and produces an additional shift of the minima 
in the spectrum to
lower photon energies (Fig.~\ref{jpsipsi}c).  This results in significant effects, 
especially, 
if one analyses the data as a function of the mass number A at different 
energies to extract A-dependence of the $J/\psi N$ cross
sections.
It is seen from Fig.\ref{adpsi} that the oscillating form factor significantly
influences the A-dependence that complicates  extracting of the genuine 
$J/\psi N$ and $\psi' N$ cross section from such analysis. 
 The oscillation of the cross section with energy 
can be used to measure in a new way 
the elementary
charmonium photoproduction amplitudes as well as  
the charmonium-nucleon amplitudes using the coherent charmonium
photoproduction off light nuclei\cite{fgszloch}. 
 In particular, at the photon energies $\omega_{\gamma}\approx 0.13R_A
m_{\psi'}^2$ (Fig.~\ref{jpsipsi}d), the main
contribution to the cross section originates from the
nondiagonal rescattering.
As a result, one can extract from the data the nondiagonal elementary
$J/\psi N\to \psi' N$ amplitude by measuring the ratio
of the  $\psi'$
 and $J/\psi$ yields at zero production angle, because other inputs to this
ratio such as the elementary $\gamma N\to J/\psi N$ amplitude are fixed from
the $J/\psi$ production.
 Moreover, since other parameters 
 enter both in the numerator and denominator, the
major uncertainties are canceled out. On the other side,
the energy dependence of this ratio 
should originate primarily due to the contribution of the direct $\psi'$
production. Thus, one would be able to determine the $\gamma N\to \psi' N$
amplitude from the measurement of the  $\psi'$-to-$J/\psi$ ratio.
We want to emphasize that the suggested procedure for extracting the
nondiagonal amplitude and amplitudes of direct $J/\psi$ and $\psi'$
photoproduction from the nuclear measurements is practically model
independent.

It is much more difficult to determine 
the diagonal $J/\psi N\to J/\psi N$ and $\psi' N\to 
\psi' N$ amplitudes, which are relevant for the suppression of the charmonium 
yield in heavy ion collisions.

A naive diagonal VDM with $\sigma_{tot}(J/\psi N)$ based on the 
SLAC data~\cite{slac1} leads to 
 a rather significant
suppression of the  $J/\psi$  yield: 
 $\approx 10\%\div 15\%$ for light nuclei, and 
 $\approx 30\%\div 40\%$ for heavy nuclei. However,    
we find~\cite{fgszch} a strong compensation of the suppression due to the
contribution of the nondiagonal transitions. As a result we find that
overall the suppression does not exceed $5\%\div 10\%$ for all nuclei along
the periodical table.  Hence, an extraction of the diagonal amplitudes from
the measured cross sections would require a comparison of high precision
data with very accurate theoretical calculations including the nondiagonal
transitions. Another strategy is possible
if the elementary $\gamma
N\to J/\psi N$, and $\gamma N\to \psi' N$ photoproduction amplitudes,
as well as the nondiagonal amplitude $J/\psi N\leftrightarrow \psi' N$
would be reliably determined from the medium  energy data.
In this case  it would be
possible to determine the imaginary parts of the forward diagonal amplitudes
from the GVDM equations~(\ref{gvdm})
\begin{eqnarray}   
\Im f_{J/\psi N\to J/\psi N}= \frac {f_{J/\psi }} {e} \Im  
f_{\gamma N \to J/\psi N}-   
\frac {f_{J/\psi }} {f_{\psi'}} \Im f_{\psi' N\to J/\psi N}\quad,    
\label{gvdm3}   
\\   
\Im f_{\psi' N\to \psi' N}=   
\frac {f_{\psi'}} {e} \Im f_{\gamma N\to \psi' N} -   
\frac {f_{\psi'}} {f_{J/\psi }} \Im f_{J/\psi N\to \psi' N}\quad.   
\label{gvdm4}   
\end{eqnarray}   
   The suggested procedure allows 
to determine all elementary
amplitudes with a reasonable precision from the  measurements
of  
$J/\psi$  and $\psi'$ 
 photoproduction 
off the light nucleus at the medium photon energies.
The measurement of the coherent charmonium photoproduction off nuclei
in this energy region are planned at SLAC~\cite{e160}. The measurements 
of coherent production off  heavy nuclei and of the quasielastic production
with parameters of GVDM fixed from the analysis of light nuclei
would provide a critical test of the model.

The main limitation of the suggested procedure is the restriction of the
hadronic basis to the two lowest 1S, 2S charmonium states with the photon
quantum numbers: $J/\psi$ , $\psi'$. 
This is one of the key approximations  in the discussed  approach.
It seems quite reasonable 
since in the coherent production of vector meson
states off nuclei at moderate energies contribution of high mass states is more 
suppressed by the target form
factor. However, we neglected by  $\psi''$ which is nearly degenerate in mass with $\psi'$: $\Delta M = 91$ MeV.
   The properties of these two states 
 are  described well in the 
charmonium model~\cite{Eichten,Richard}. In this model
$\psi''$-meson  is described as  a ${}^3D_1$ state   
 with a small admixture of the $S$ wave, while  $\psi'$-meson
has a small $D$-wave admixture. Namely
\begin{eqnarray}
\left|\psi'\right>&=&\cos \theta  \left|2S\right> + \sin\theta 
\left|1D\right>,\cr
\left|\psi''\right>&=&\cos \theta  \left|1D\right> - \sin\theta 
\left|2S\right>\quad.
\end{eqnarray}
Since only the S-wave contributes to the decay 
of $\psi$ states into $e^+e^-$ 
(at least in the non-relativistic charmonium models)
the value  $\theta=19^o \pm 2^o$ can be determined from the 
data on the $e^+e^-$ decay widths  
$\Gamma (\psi'\rightarrow e^{+}e^{-})=2.19\pm0.15$ KeV
 and the $\Gamma (\psi"\rightarrow e^{+}e^{-})=0.26\pm 0.04$ KeV.
Due to the small difference of masses between the 
$\psi ^\prime$ and $\psi ^{\prime \prime}$
mesons the produced S-wave $c\bar c $-state
does not loose coherence while going through the
media at any conceivable energies. The soft  interactions cannot
transform the $S$-state to $D$-state with any significant probability. 
In the soft QCD processes data show that the cross sections
of exclusive nondiagonal transitions are negligible for the forward angle 
scattering. The same conclusion is valid in the PQCD model for the
charm dipole-nucleon interactions.
Thus, it is  more appropriate to use the $1S$ - $2S$ 
basis for description of the propagation of $c\bar c$ through the nucleus. 
The only resulting change is  an increase of $f_{2S}$ 
by a factor of $1/\cos\theta$ as compared to $f_{\psi'}$
which is within the uncertainties of the model. Since the $2S$ state ultimately
transforms into $\psi'$ and $\psi"$ we predict ~\cite{fgszpsip}: ~
\footnote{If the s-wave mechanism 
dominates in the charmonium production in other hard processes the $\psi"/\psi'$ ratio would
be a universal number. Since $\psi"$ can be easily observed via its characteristic
$D\bar D$ decays in a number of the 
lepton, hadron and nucleus induced processes the study of the discussed process can 
provide a new 
way to probing dynamics of charmonium production in various reactions.}
\begin{equation}
{\sigma(\psi'')\over\sigma(\psi')}=\tan^2 (\theta) \approx 0.1\quad.
\end{equation} 

Influence of 
the higher mass resonances is expected to be even weaker - the constants
$1/f_{V}$ relevant for the transition of a photon to a charmonium state 
$V$ rapidly decrease with the resonance mass. This is because the radius of 
a bound state, $r_V$, is increasing with the mass of the resonance and
therefore the probability of the small size configuration being $\propto
1/r_{V}^3$ is decreasing with an increase of mass (for fixed S,L).  
Besides, the asymptotic freedom in QCD dictates decreasing of the coupling
constant relevant for the behavior of the charmonium wave function at small
relative distances.  Experimentally one finds from the data on the leptonic
decay widths that $1/f_{V}$ drops by a large factor with increasing mass. 
An additional suppression arises due to the weakening of the soft
exclusive nondiagonal $V N\leftrightarrow V' N$ amplitudes between
states with the different number of nodes. Hence,
it seems possible to  determine the
imaginary parts of diagonal rescattering amplitudes from analysis of the data
using
 the GVDM equations.
  Since in the medium energy domain the energy
dependence of soft rescattering amplitudes is well reproduced by a
factor $s^{0.08}$, one can determine  the real parts of the amplitudes
using the 
Gribov-Migdal relation (Eq.~\ref{GM}).
  
The above analyses demonstrate that the 
relative importance of the non-diagonal transitions increases
 an increase of the quark mass.
 It
would be of interest to investigate experimentally the case of
strangeness production. It  is likely to correspond to $\epsilon $
substantially larger than  $\epsilon_{\rho}\sim 0.2$ but much smaller
than $\epsilon_{J/\psi}\sim 1.7$.

\section{Onset of perturbative color opacity at small x and onium
  coherent photoproduction.}

Interaction of small size color singlet objects with hadrons is one
of the most actively studied issues in high-energy QCD. The QCD 
factorization theorem for exclusive meson 
electroproduction at large $Q^2$, and   $J/\psi,\Upsilon$ photoproduction 
\cite{CFS,BFGMS} allows to evaluate the amplitude   of the production 
of a vector meson by a longitudinally polarized photon 
$\gamma_L + T \to V + T$ through the convolution of the wave function
of the meson at the zero transverse separation, hard interaction block
and the generalized (skewed) parton density
\footnote{Proportionality of the hard
  diffractive amplitudes to the gluon density of the nucleon
was discussed for hard pp diffraction in \cite{FS89}, 
and for $ J/\psi$ production \cite{Ryskin} in the BFKL approximation
and for the pion  diffraction into two jets \cite{FMS} in the leading
log $Q^2$ approximation \cite{FMS}.}.  The  LT approximation differs
strongly from the expectations based on Glauber model approaches and 
on the two gluon exchange models, because it accounts for the
dominance of the electroproduction of spatially small quark-gluon wave package
and its space-time evolution which 
leads to  formation of a softer gluon field. The latter effect
results effectively  in an  increase of the size of the  dipole with
increase of the energy.

In perturbative QCD  (similar to QED) the total cross section of the 
interaction of small systems with hadrons is proportional to the area
occupied by color within projectile hadron \cite{Low} leading to the 
expectation of the 
color transparency phenomenon for various hard processes with nuclei.
The cross sections of incoherent processes are expected to be proportional 
to the number of nucleons in the nuclei, while  the  coherent 
amplitude is proportional to number of nucleons times the 
nuclear form factor. Possibility to approximate projectile heavy quarkonium 
as colorless dipole of heavy quarks can be formally derived from QCD within 
the limit when mass of heavy quark $m_{Q}\rightarrow \infty$ but 
$x=4m_{Q}^2/s$ is fixed and not extremely small \cite{FKS}. In this 
kinematics the size of heavy quarkonium is sufficiently small to justify
applicability of PQCD.

 For practical purposes the  crucial question is at what $Q^2$ 
squeezing becomes effective. Probably the most sensitive indicator 
is the $t$-dependence of the meson production. The current HERA data
are consistent with the prediction of \cite{BFGMS,FKS}
that the slopes of the $\rho$ and $J/\psi$ production amplitudes 
should converge to the same value. This indicates that at small ${\it x}$
configurations much smaller than average configurations 
in light mesons ($d\sim 0.6 fm$) dominate for $\rho$-meson
production at  $Q^2\ge 5$ GeV$^2$, while 
the $J/\psi$ production is dominated by interaction in small size configurations for all $Q^2$.
Therefore,  one expects the regime of color transparency
for ${\it x}\ge 0.03$ where the gluon shadowing is very small/absent.

Recently the color transparency (CT) phenomenon was observed
at FNAL by E791 experiment~\cite{E791} which studied  the coherent
process of dissociation of a 500 GeV  pion
into two jets off the nuclei. The measurement has confirmed a number of
predictions of \cite{FMS} including the A-dependence, and the transverse
and longitudinal momentum distributions of the jets.
Previously the color transparency type behavior of the cross section
was observed also in the coherent $J/\psi $ photoproduction
at $\left<E_{\gamma}\right>=120 GeV$ \cite{Sokoloff}.

A natural question is whether the color transparency will hold
for arbitrary high energies? Two phenomena are expected to work
against CT at high energies.
One is the
LT gluon shadowing. 
{}There are theoretical expectations (see discussion below) 
supported to some extent by the current   analyses of the data on  DIS
scattering off nuclei  (which do not
extend deep enough into the shadowing region) that  the  gluon distributions are 
shadowed
in nuclei as compared to  the nucleon: $G_A(x,Q^2)/A G_N(x,Q^2) < 1$.
This obviously should lead to a gradual
but calculable in QCD
disappearance of color transparency \cite{BFGMS,FMS}, and 
 to onset of a new regime,
which we refer to as {\it the  color opacity regime} 
(one can think of this regime also as a regime of
 generalized color transparency since a small $q\bar q$ dipole still couples to the gluon
field of the target through a two gluon attachment and the 
amplitude is  proportional to the generalized gluon
density of the nucleus).
Another
mechanism for the violation of CT at high energies is the increase of the
small dipole-nucleon cross section with energy $\propto G_{N}(x,Q^2)$.
For sufficiently large energies this cross section becomes comparable to 
the meson-nucleon cross sections.  One may expect that this would result in 
a significant  
suppression  of the hard exclusive diffractive processes
 as compared to the 
LT approximation. 
However it seems that this phenomenon is beyond the kinematics achievable
for the photoproduction of $J/\psi$-mesons in UPC of heavy ions at RHIC
($x\approx 0.015,Q^2_{eff} \approx 4 GeV^2$)
but could  be important at heavy ion UPC at LHC.

Hence, a systematic study of the onium production in the coherent
scattering off nuclei at collider energies will be very interesting.
One should emphasize here that with decrease of the size of the onium 
the eikonal (higher twist) contributions 
die out quickly (provided $x$ is kept fixed). 
In particular, for the $\Upsilon $ case one probes 
nuclear gluon fields at the transverse scale of the order of 0.1 fm or
$Q^2_{eff} \sim 40 GeV^2$.  The $J/\psi$ case is  closer to the border line
between the perturbative and nonperturbative domains. As a result the
a nonperturbative region appears to give a significant contribution to
the production amplitude \cite{FMS2000}.

\begin{figure}
\centering
        \leavevmode
        \epsfxsize=.50\hsize
       \epsffile{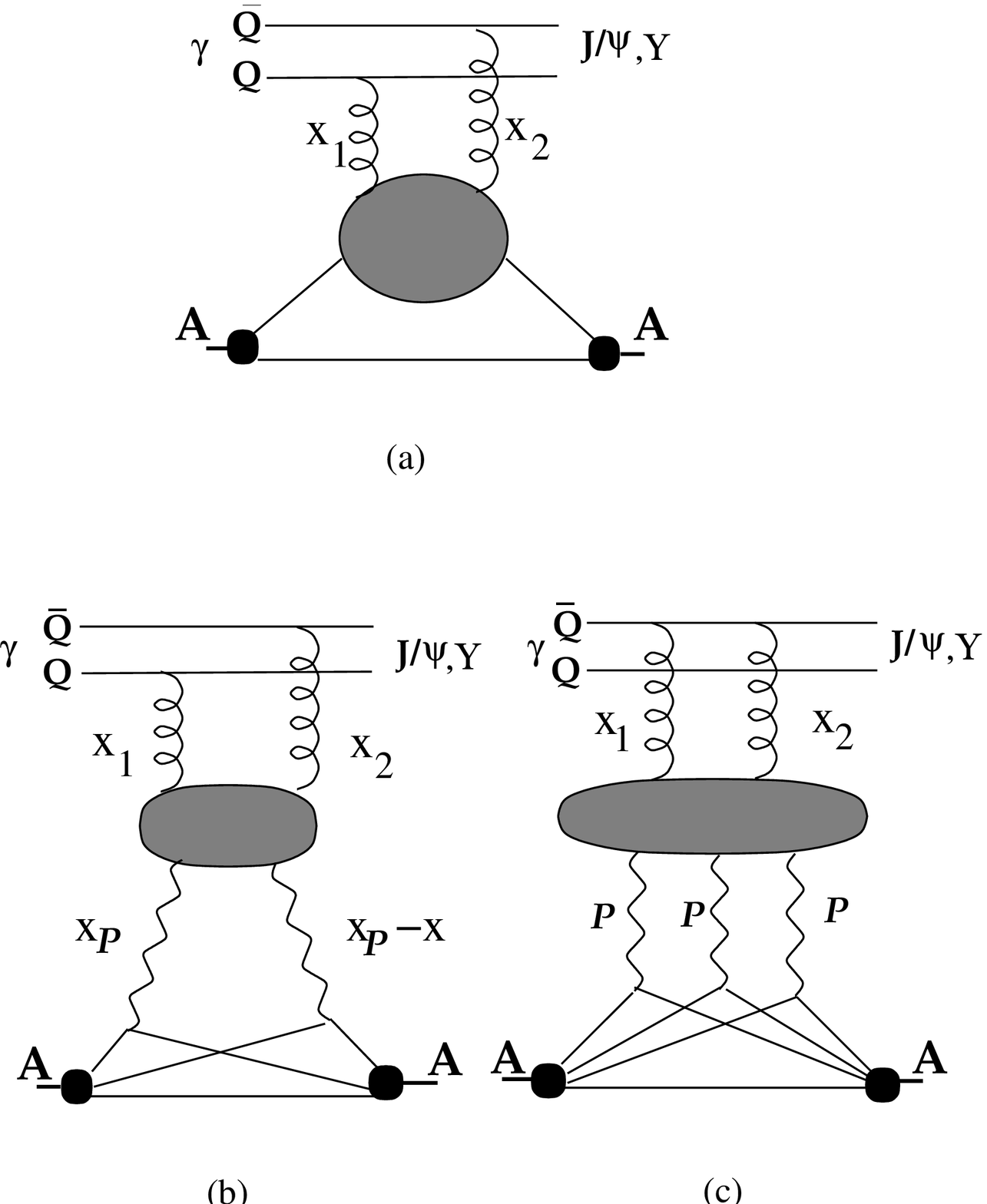}
\vspace*{1cm}
\caption{Leading twist diagrams for the   production of
quarkonium off nucleus.}
\label{ltdiag}
\end{figure}

Let us discuss the photoproduction amplitude
$\gamma + A\to J/\psi(\Upsilon) +A$ in more details.
We are interested here in the $W_{\gamma p}$ range which 
can be probed at LHC. This region corresponds to rather small values of ${\it x}$.
In this situation interaction of $q\bar q$ pair, which in the final state
forms a quarkonium state is still rather far from the BBL. Hence 
 the key problem in the theoretical treatment of the process is taking 
into account  the nuclear shadowing.
A number of  mechanisms of coherent  interactions with several   were
suggested    for this process. We focus here on the
 {\bf the leading twist mechanism of shadowing} which is presented
by diagrams in Fig.\ref{ltdiag}.
There exists  qualitative difference between
the mechanism of interaction of a small dipole with several  nucleons
and the case of a similar interaction of an ordinary hadron.
Let us for example consider interaction with two nucleons.
The leading twist contribution is described by the diagrams
where two gluons are attached to the
dipole. To ensure that nucleus remains intact in such a process, the 
 color singlet  lines should be attached to both nucleons.
These diagrams (especially the one of Fig.~\ref{ltdiag}b) are closely related
 to the diagrams describing 
the gluon diffractive parton densities (which are measured
at HERA),   and hence to the
similar diagrams for the gluon nuclear shadowing \cite{FS99}.

The amplitude of  high energy coherent 
heavy onium photoproduction is proportional to the generalized gluon density of the target,
$G_T(x_1,x_2,t,Q_{eff}^2)$,
which depends on the light-cone fractions  $x_1$ and $x_2$ of gluons attached to the 
quark loop. They satisfy a relation:
\begin{equation}
x_1-x_2={m^2_V\over s}\equiv x
\end{equation}
If the  quark Fermi motion and binding effects were negligible,
$x_2\ll x_1$. The resolution scale $Q_{eff}^2\ge m_q^2$, where $m_q$ is
the mass of the heavy quark.
Numerical estimates for the photoproduction of $J/\psi$ give
 $Q^2_{eff} \sim 3- 4 GeV^2 $\cite{FKS,FMS2000}
reflecting a relatively small mass of c-quark and 
indicating that this process is on the verge between nonperturbative 
and perturbative regimes.
On the contrary, the  mass of the beauty quark 
is huge on the scale of soft QCD. In this case hard physics 
dominates, attachments of more than two gluons to $b\bar b$ are negligible 
and the QCD factorization theorem provides a reliable  
tool for  description of the  $\Upsilon$ production.
This is especially true for the ratio of the cross sections of $\Upsilon$ production from
different targets since the higher twist effects due to the 
overlapping integral of the $b\bar b$ component of 
the photon and $\Upsilon$ cancelled out  in the ratio.
 As a result, in the leading twist shadowing 
approximation the cross section of the process 
$\gamma A\rightarrow \Upsilon A$ is proportional to the squared nuclear gluon
density distribution  and can be written in the form  
\begin{equation}
\sigma_{\gamma A\rightarrow V A}(s)=  
{d \sigma_{\gamma N \to V N}(s,t_{min})\over dt}
\Biggl [\frac {G_{A}(x_1,x_2,Q_{eff}^2,t=0)}
{AG_{N}(x_x,x_2,Q_{eff}^2,t=0)}\Biggr ]^2
\int \limits_{-\infty}^{t_{min}} dt
{\left|
\int d^2bdz e^{i{\vec q_t}\cdot {\vec b}}
e^{-q_{l}z}\rho ({\vec b},z)
\right|
}^2.
\label{phocs}
\end{equation}

  Numerical estimates using realistic potential model wave
functions indicate that for  $J/\psi$, $x_1\sim 1.5 x, x_2\sim x/2$
\cite{FMS2000}, and for $\Upsilon$, $x_2/x_1\sim 0.1$ \cite{fmsupsilon}.
Modeling of the generalized parton distributions at moderate $Q^2$ 
suggests that to a good
approximation $G(x_1,x_2,t=0) $ can be approximated by the gluon density
at $x=(x_1+x_2)/2$ \cite{BFGMS,Rad}. For large $Q^2$ and small $x$ GPDs
are dominated by the evolution from  $x_i(init) \gg
x_i$. Since the evolution conserves $x_1-x_2$, effect of skewedness is
determined primarily by the evolution from nearly diagonal  distributions, see 
Ref.~\cite{MF}
and references therein.
In the case of the $\Upsilon $ production it increases the cross section
by a factor  $\sim 2 $ 
\cite{fmsupsilon,Martin:1999rn} and, potentially, this could even 
obscure
the connection of the discussed effect with the shadowing of the nuclear gluon
densities. However, the  analysis of  \cite{fgsrev} shows that 
the ratio of GPD on a nucleus and on a nucleon at $t=0$ 
is a weak function of $x_2$,
slowly dropping from its diagonal value ($x_2=x_1$) with the decrease
of $x_2$. Overall this observation is in a agreement with the general
trend mentioned above that it is more appropriate to do comparison of
diagonal and non-diagonal cases at $x=(x_1+x_2)/2$.

In the case of double scattering contribution (Fig.~\ref{ltdiag}b) there is
another way to address the question of the accuracy of 
the substitution of the ratio of generalized gluon densities by the
ratio of diagonal parton densities  at the normalization scale.
We notice that the amplitude corresponding
to  Fig.~\ref{ltdiag}b is expressed through
the nondiagonal matrix element of the diffractive 
distribution function, $\tilde{g}^D(x_{\Pomeron},Q^2,x_1,x,t)$, 
which is
 an analog of generalized parton distribution. In the 
diagonal limit of $x_1=x_2$ it 
coincides with the diffractive gluon  distribution.
It depends on the light-cone 
fraction which nucleon lost in  $\left|in \right>$ and 
$\left<out\right|$ states :  $x_{\Pomeron}$
$x_{\Pomeron}-x$ respectively,  $\beta_{in}=x_1/x_{\Pomeron},
\beta_{out}=(x_1-x)/(x_{\Pomeron}-x)$,  and $t,Q^2$.
If we make a natural assumption that 
\begin{equation}
\tilde{g}^D(x_1,x,x_{\Pomeron},Q_0^2,t)=
\sqrt{g^D(\beta_{in},Q^2_0,x_{\Pomeron},t)
g^D(\beta_{out},Q^2_0,x_{\Pomeron}-x,t)},
 \end{equation}
we find that numerically in the kinematics we discuss the resulting
skewedness effects are small  as compared to the uncertainties
in the input gluon diagonal  diffractive PDFs. 

Hence in the following we will approximate the ratio of generalized 
gluon densities in the nucleus and nucleon by the ratio of 
the gluon densities in nucleus and nucleon at $x=m_V^2/s$. In the case of 
$\Upsilon$ use of $\tilde x \sim x/2$ maybe more appropriate. 
This would lead to slightly larger shadowing effect. 

It was demonstrated in \cite{FS99} that one can
express the  quark and gluon nuclear shadowing
for the interaction with two nucleons in a model independent
way through the corresponding diffractive parton densities using
the Gribov theory of inelastic shadowing~\cite{Gribovinel} and
the QCD factorization theorem for the hard diffraction
\cite{Collins}.  An important discovery of HERA is that
hard diffraction  is indeed dominated by the leading twist contribution and
gluons play a very important role in the diffraction(this is loosely
referred to as gluon dominance of the Pomeron). Analysis of the
HERA diffractive data indicates that in the gluon induced processes
probability of the diffraction
is much larger than in the quark induced processes
\cite{FS99}. The recent H1 data on diffractive dijet production
\cite{H1} provide an additional confirmation of this observation.
Large probability of diffraction in the gluon induced hard processes
could  be understood in the s-channel language as formation of
color octet dipoles  of rather
large sizes which can diffractively  scatter with a quite  large cross
section. The strength of this interaction can be quantified using
optical theorem and introducing
\begin{equation}
\sigma_{eff}^g=\frac {16 \pi} {\sigma_{tot}(x,Q^2)} 
{d \sigma_{diff}(x, Q^2,t_{min})\over dt}
=\frac {16\pi } {(1+{\eta}^2)G_N(x,{Q^2})}
\int \limits_{\it x}^{{\it x}_{\Pomeron}^{0}}
d{\it x}_{\Pomeron}
g^{D}_{N}(\frac {\it x} {{\it x}_{\Pomeron}},{\it x}_{\Pomeron},
Q^2,t_{min}),
\label{sigef}
\end{equation}
for the hard process of scattering of a virtual photon off the gluon
field of the nucleon. 
Here $\eta$ is the
ratio of the real to imaginary parts of the elementary diffractive amplitude,
$Q^2$ is the momentum scale determining virtuality of the gluons,
${\it x}_{\Pomeron}$ is the momentum fraction of the pomeron 
with the corresponding cut-off scale  ${\it x}_{\Pomeron}^{0}=0.03$,
and  $g^{D}_{N}(\frac {\it x} {{\it x}_{\Pomeron}},{\it x}_{\Pomeron},
Q^2,t_{min})$ is the diffractive gluon density distribution 
of nucleon which is
known from the H1 analysis of the diffractive data at the scale $Q^2\approx 4$ GeV$^2$. 

An important feature of this mechanism of
coherent interaction is that it is practically absent for
$x\ge 0.02\div 0.03$ and may rather quickly become important with
decrease of $x$.

\begin{figure}
\centering
        \leavevmode
        \epsfxsize=.80\hsize
       \epsffile{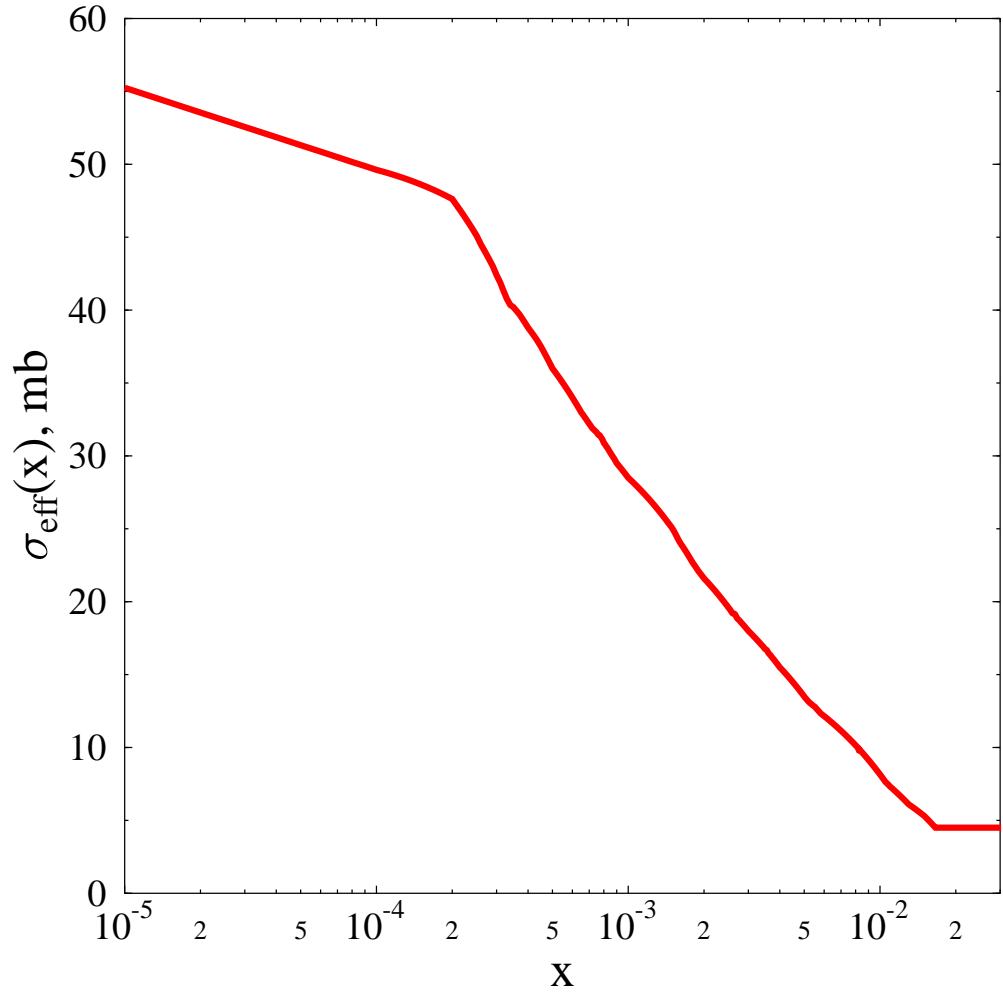}
\vspace*{1cm}
\caption{
The effective cross section shadowing in the gluon channel,
$\sigma_{eff}(x)$, at $Q^2=4$ GeV$^2$ as a function of the
Bjorken x for H1 parameterizations of the gluon diffractive density.}
\label{figsigef}
\end{figure}

We calculated the  ratio of the gluon density distributions in 
Eq.~\ref{phocs} using the leading twist shadowing model~\cite{FS99}
(for the details of calculations see Ref.\cite{guide}). As the  first step
the nuclear gluon density distribution with account of  the leading
twist shadowing is calculated at the starting evolution scale 
$Q_{0}^2=4$ GeV$^2$ 
\begin{eqnarray}
\nonumber
 {G_{A}({\it x},Q_{0}^2)}= {AG_{N}({\it x},Q_{0}^2)}-
8\pi \Re \biggl [
{\frac {(1-i\eta )^2} {1+{\eta}^2}}
\int d\,^2\,b\, \int \limits_{-\infty}^{\infty} dz_1
\int \limits_{z_1}^{\infty}dz_2\int \limits_{\it x}^{{\it x}_{\Pomeron}^{0}}
d{\it x}_{\Pomeron}\\
g^{D}_{N}(\frac {\it x} {{\it x}_{\Pomeron}},{\it x}_{\Pomeron},
Q_{0}^2,t_{min})\rho({\vec b},z_1)\rho({\vec b},z_2)e^{i{\it x}_{\Pomeron}
m_N(z_1-z_2)}\,e^{-{\frac {1} {2}}\sigma_{eff}(x,Q_{0}^2)(1-i\eta )\int 
\limits_{z_1}^{z_2} dz\rho({\vec b},z)}
\biggr ].
\label{ga}
\end{eqnarray}
As an input we used the H1 parameterization of
 $g^{D}_{N}(\frac {\it x} {{\it x}_{\Pomeron}},{\it x}_{\Pomeron},Q_{0}^2,t_{min})$.
The  effective cross section $\sigma_{eff}(x,Q_{0}^2)$ accounts for the
elastic rescattering of the produced diffractive state off the nuclear nucleon
and is determined by Eq. \ref{sigef}. Numerically it is
very large at the starting scale of the evolution, see
 Fig.~\ref{figsigef}, and corresponds to the probability of the gluon
induced diffraction close to 50\%.
This indicates that at the initial scale of the evolution interactions
 in the gluon sector are close to the BBL for $x\le 10^{-3}$
 for the nucleon case,  and even more so for the nuclei, where similar
 regime should hold for a larger range of the impact parameters.

Note that  the double scattering term 
in  Eq.~\ref{ga} for the nuclear parton densities 
satisfies QCD evolution,
 but 
the higher order terms do  not. 
That is if we use different starting scale of evolution 
we would obtain different results of $G(x,Q^2)$.
The reason is that the terms $\propto \sigma_{eff}^n, n\ge 2$ are
sensitive to degree of fluctuations in the cross sections of
interaction of diffracting states. These fluctuations increase with
increase of $Q^2$. This effect is automatically 
included in the QCD evolution, and it leads to violation of the Glauber - like 
structure of the expression for the shadowing at $Q^2>Q_0^2$.
Approximation for the $n\ge 3$ rescattering of Eq.~\ref{ga}
corresponds to an assumption that
 fluctuations are small at $Q_0^2 $
scale since this scale is close enough to the scale of the soft
interactions, see discussion in \cite{FS99}.
Thus,  at the second step of the calculation we use NLO QCD evolution
equations
to calculate the shadowing at larger $Q^2$ using the calculation
at $Q^2_0$ as a boundary condition. In this way we also take
into account the contribution of the gluon enhancement at $x\sim 0.1$
which influences the shadowing at larger $Q^2$.
Note here that 
proximity to the BBL in the gluon sector which is reflected in a large value of
$\sigma_{eff}$ may result in
corrections to the LT evolution which require further studies.

\begin{figure}
\centering
\epsfxsize=0.8\hsize
   \epsffile{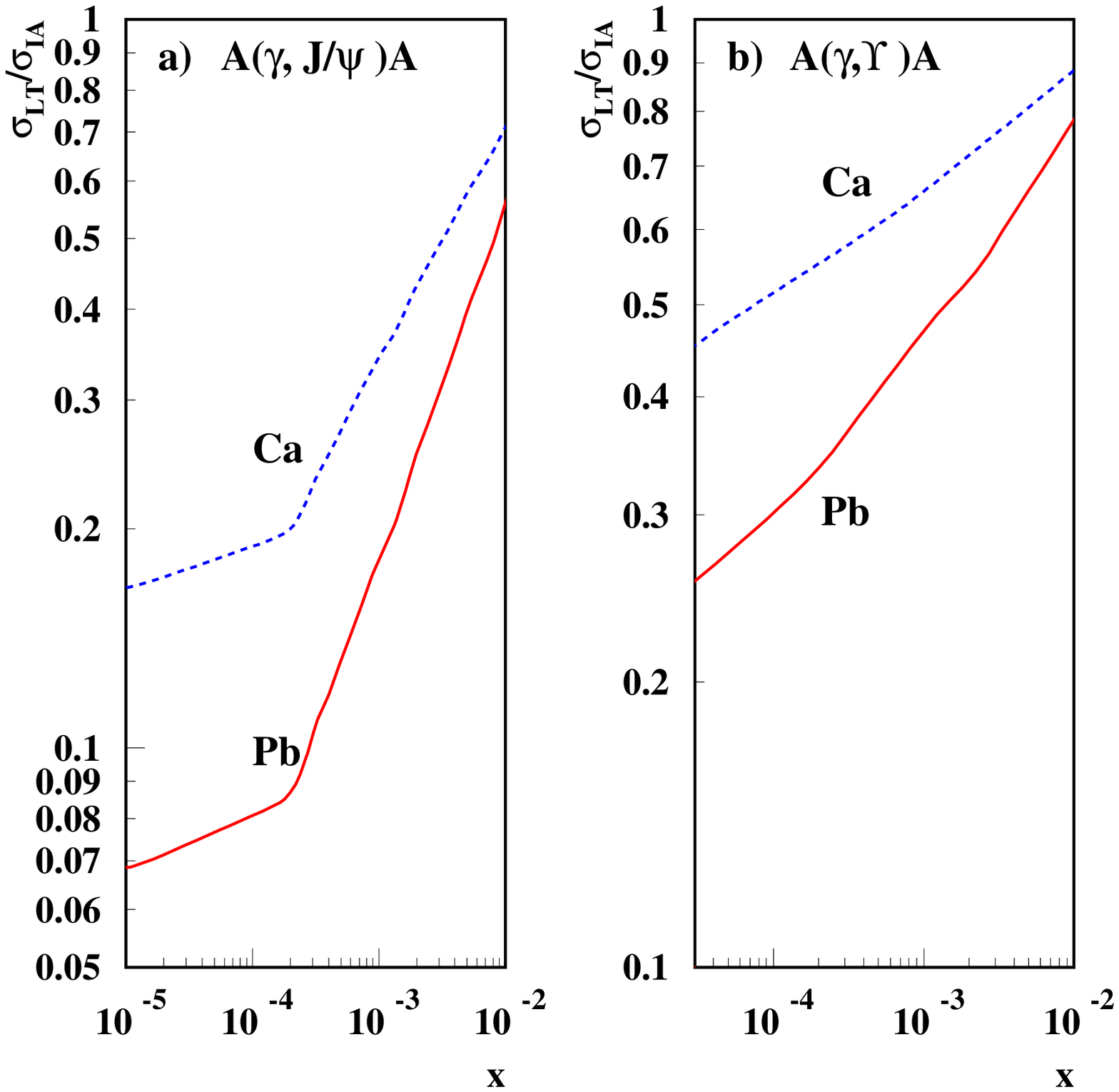}
\caption{The ${\it x}$-dependence of the ratio 
of $J/\psi$ and $\Upsilon$ production off Ca and Pb in Glauber model to that
in Impulse Approximation(IA).
Calculation with the H1 parameterization\cite{H1} of the diffractive PDF.}
\label{fshad}
\end{figure}

First we calculate the ratios of the cross sections of coherent
 photoproduction of $J/\psi$ and $\Upsilon$ off nuclei and nucleon
(Fig.\ \ref{fshad}).
Such  ratios do not  depend on the uncertainties of the 
elementary cross sections and provide a sensitive test of the role
 of the LT shadowing effects.
In a case of the $J/\psi$  photoproduction the gluon
virtuality scale 
is 3-4 GeV$^2$ with a significant fraction of the amplitude
due  to smaller virtualities \cite{FKS,FMS2000}.  Hence we will take
the gluon shadowing in the leading twist at $Q^2=4 GeV^2$. Taking a smaller
value of $Q^2$ would result in even  larger shadowing effect.
In  calculations for the $\Upsilon$ we
take $Q_{eff}^2=40 GeV^2$, though the result is not very sensitive to
precise value of $Q^2_{eff} $ since the scaling violation for the 
gluon shadowing for these $Q^2$ is rather small. We find that in spite 
of a small size of $\Upsilon$, which essentially precludes higher twist
 shadowing  effects up to very small x, the perturbative color opacity effect
is quite appreciable. It is worth noting that the effective cross
 section of the rescattering in the eikonal model is determined by the
 cross section of dipole -nucleon interactions at the distances $d\sim
 0.25 -0.3 fm$ in the $J/\psi$ case and $\sim 0.1 fm $ in the
 $\Upsilon$ case. These cross sections are $\sim   10-15 mb$
for $J/\psi$ and $\sim 3 mb$ for $\Upsilon$ case  for $x\sim 10^{-4}$
and a factor of 1.5-2 smaller for $x\sim 10^{-3}$, see Fig. 13 in
 Ref.~\cite{FMS2000}. They are much smaller than the cross sections
which enter into calculation of the gluon shadowing in the LT
 mechanism.

\begin{figure}
\centering
\epsfxsize=0.8\hsize
   \epsffile{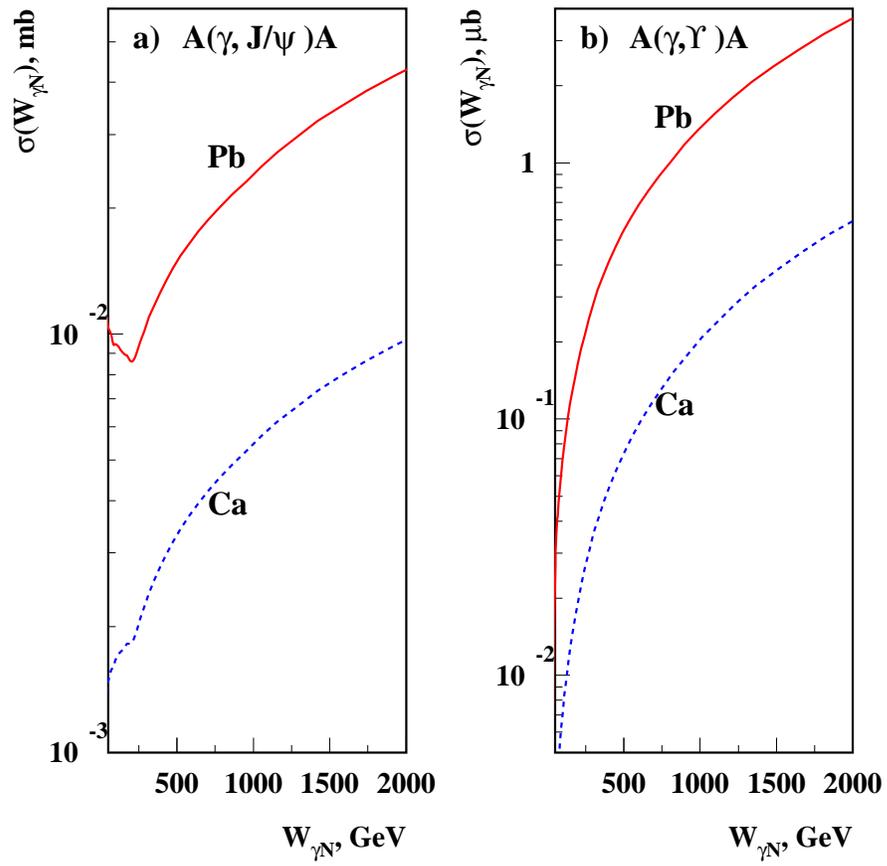}
\caption{The energy dependence of the coherent $J/\psi$ and $\Upsilon$
photoproduction off Ca and Pb in the LT approximation.}
\label{onics}
\end{figure}

We also estimated the absolute cross section of onium photoproduction
for a wide range of the photon energies. This is  of interest
for the planned measurements of the onium production
 in the ultraperipheral heavy ion collisions at LHC.
The dependence of the momentum-integrated cross sections on the energy
$W_{\gamma N}=\sqrt{s}$  is presented in Fig.~\ref{onics} . 
In the case of $J/\psi$ production 
 the calculations are  pretty straightforward as the accurate data
are available from HERA. The situation 
 is more complicated 
in the case of the 
photoproduction of $\Upsilon$. So far the information about the elementary 
$\gamma +N\to \Upsilon +N$ cross section  is very limited.
There is the only ZEUS and H1 data on total
cross section for the average energy of  $\sqrt {s} \approx 100 GeV$. 
Thus, 
to calculate the forward photoproduction cross section we used a
simple parameterization:
\begin{equation}
 {d \sigma_{\gamma N\to VN}(s,t)\over dt}=10^{-4}
B_{\Upsilon}\left({s\over s_0}\right)^{0.85}\cdot exp(B_{\Upsilon}t),
\label{eq:cs}
\end{equation}
where $s_0=6400$ GeV$^2$, the slope parameter $B_{\Upsilon}=3.5\,GeV^{-2}$ is fixed 
basing on the analysis
of the two gluon form factor in Ref.~\cite{Frankfurt:2002ka},
and the energy dependence follows from the calculations of
 Ref.~\cite{fmsupsilon}
of the photoproduction of $\Upsilon$ in the leading $\log Q^2$ approximation 
with an account for the skewedness of the partonic density distributions.
This elementary cross section is normalized so that the total
cross section is in $\mu b$.

\section{LARGE MASS DIFFRACTION IN THE LEADING TWIST LIMIT}

It is well known that inelastic diffraction at small $t$ gives information 
on the fluctuations of strength in the projectile-target interactions
\cite{GW,FP}. Application of this logic to the hadron scattering off nuclei have allowed to
explain  the A-dependence of  two  measured diffractive channels in
$pA,\pi A$ scattering
assuming that it coincides with the A-dependence of the total 
cross section of the inelastic
diffraction, 
and the absolute total 
cross section of pA diffraction 
(the data exist for two nuclei only) at the energies 200-400 GeV,
 see \cite{fgsrev} for the review and references.
With an increase of energy, the total
 cross section of $NN$ interaction increases and 
fluctuations in the elementary amplitude lead to much smaller fluctuations of the absorption
in the scattering off a heavy nuclei.  As a result, one can expect 
%and to
much weaker A-dependence
of the diffractive cross section~\cite{FMSdiff}, in particular,
$\sigma_{diff}(p+A\to X+ A)\propto A^{0.25}$ at LHC energies
\cite{felix} as compared to $\sigma_{diff}(p+A\to X+ A)\propto
A^{0.7}$ at fixed target energies. 
For large produced masses we can also understand this suppression using 
the $t$ channel picture of the Pomeron exchanges as due to
the stronger screening of the triple Pomeron exchange, 
for review and references see \cite{Kaidalov}.

Diffraction in deep inelastic scattering
corresponds to the transition of the (virtual) photon
into its hadronic components leaving the nucleus intact.
Hence it is similar to elastic hadron-nucleus scattering rather than inelastic diffractive
hadron-nucleus scattering.

It was demonstrated recently that in the UPC collisions at LHC it would be possible 
to study nuclear parton densities using hard charm and beauty production in
$\gamma + A$ interactions~\cite{KleinVogt}. Naturally one can also use these 
and similar processes to measure  diffractive parton densities of nuclei.
Since these quantities in the leading twist satisfy the factorization theorem we can 
analyze them on the basis of the analysis 
of  the diffraction in DIS.

There is a deep connection between shadowing and phenomenon of 
diffractive scattering off nuclei.
The simplest way to investigate this connection is to apply 
the AGK cutting rules~\cite{AGK}. Several processes contribute to 
diffraction on nuclei: (i) Coherent diffraction in which the nucleus 
remains intact, (ii) Break-up of the nucleus without production of hadrons in
the nucleus fragmentation region, 
(iii) Rapidity gap events with hadron production in the 
nucleus fragmentation region. In Ref. \cite{FSAGK} we found that for 
$x\leq 3\cdot 10^{-3}, Q^2\ge 4$ GeV$^2$, the fraction of the DIS events 
with rapidity gaps reaches the value of about 30-40\% for heavy 
nuclei, with a fraction of the events of type (iii) rapidly dropping with $A$.

\begin{figure}
\centering
\epsfxsize=0.8\hsize
   \epsffile{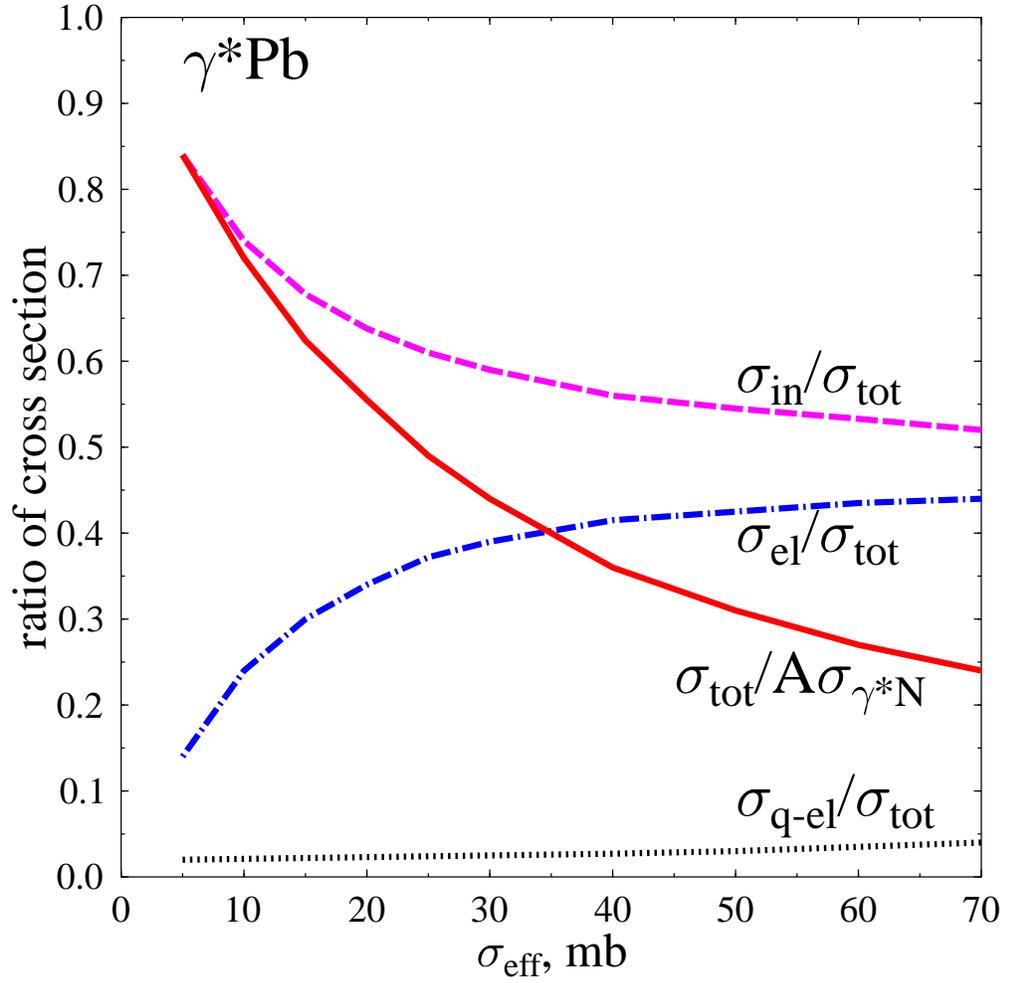}
\caption{Dependence of the partial cross sections
ratios for  the hard
 processes on the effective cross section defined in Eq.21 compared
 with the shadowing of the inclusive hard process.}
\label{figcsratio}
\end{figure}

We can use the information on $\sigma_{eff}$ for quarks and gluons to
estimate probability of diffraction for different hard triggers
at the resolution scale $\sim Q_0^2$.
First we consider the dependence of the fraction of the events due to 
coherent diffraction and due to the break-up of the nucleus  on the 
strength of the interaction,
$\sigma_{eff}^j$, neglecting fluctuations of the interaction strength.
We find that for the realistic values of 
$\sigma_{eff}^j$ the probability of coherent diffraction 
is quite large but increases with $\sigma_{eff}$ very slowly and does not
reach the asymptotic value of $1/2$ even for very large values of
$\sigma$ (the later feature reflects presence of a significant
diffuse edge even in heavy nuclei), see Fig.~\ref{figcsratio}. 
Thus, it is not sensitive to the fluctuations of $\sigma_{eff}$. 
We also found that the ratio of diffraction with the
nucleus break-up and with the nucleus remaining intact is small (10-20\%) 
in a wide range of nuclei,  and slowly increasing with increase of
$\sigma_{eff}$, see Fig.~\ref{figcsratio}. Hence it would require high
precision measurements to constrain the dynamics using
$\sigma_{q-el}/\sigma_{tot}$ ratios. 

Comparing the values of the fraction of the diffractive events for
quark and gluon induced processes off heavy nuclei and  proton, we find 
that the relative importance of the quark induced events is
increasing. Therefore,  the scaling violation at large $\beta$ 
for the diffractive quark distribution in nuclei will be stronger 
for nuclei than for a  proton.  Another 
interesting effect is that for heavy nuclei only genuine elastic 
components can be produced (inelastic diffraction is zero). Hence, the
soft contribution at $Q^2_0$ due to triple Pomeron exchange is strongly 
suppressed see e.g. \cite{FSAGK}. As a result,
 nuclear diffractive parton distributions
 at small $\beta$ are  strongly suppressed (by a factor $\propto A^{1/3}$)
at  $Q^2_0$ though this suppression will be less pronounced
at large $Q^2$ due to the QCD evolution.
 This will lead to breakdown of the universality
 of the $\beta$ distributions as a function of A.

%%%
Though the diffractive parton densities change rather slowly with 
$Q^2$ leading to a weak variation of the diffractive cross sections with $Q^2$
(modulus the scaling factor) the fraction of the diffractive events 
at fixed $x$ should significantly  drop with increase of $Q^2$ 
due to a large
 increase of the inclusive nucleon parton densities and decrease
 of the nuclear shadowing.

For example, let us consider 
 ultraperipheral collisions (UPC) at LHC where one 
can measure a  the process $\gamma +A \to jet_1 +jet_2 +X +A $
in the kinematics where direct photon process 
$\gamma +g \to q\bar q$ dominates.
In this case if we consider the process at say $p_t \sim 10 GeV/c$ 
corresponding to $Q^2\sim 100 GeV^2$ 
 the fraction of diffractive events will be of the order 10\%. 
%%%
The background from the strong interaction originates from glancing collisions
in which  two  nucleons interact via 
a double diffractive process $pp\to pp+ X$ where $X$ contains jets.
 Probability of the hard processes with two gaps is very small 
at collider energies - even smaller than .the probability of the
 single diffractive hard processes, see e.g. \cite{Dino}. 
Therefore, 
we expect that the background conditions
 will be at least as good in the diffractive case as in the
inclusive case considered in \cite{KleinVogt}.
Thus, it would  be pretty straightforward to extract coherent diffraction 
by simply using  anti-coincidence with the forward neutron  detector,
especially in the case of heavy nuclei, see discussion in \cite{STZ}.
As a result it would be possible to measure in 
the UPC the nuclear diffractive parton distributions 
with a high statistical 
accuracy. It is important that in difference from the diffraction
 to a vector meson it would be possible 
to determine on the event
 by event basis the energy of the photon which induced 
the reaction, 
 since the rapidity of the photon 
is close to the rapidity of two jets.
As a result it would be possible to perform the measurements for 
large rapidities (selecting the events generated by  a photon of higher of two
 energies allowed by the kinematics of production particles
in the interval of rapidities $ y_1 < y < y_2$)
and  to determine diffractive parton 
densities for pretty  small $x$.
%%%%

\section{LARGE MASS DIFFRACTION IN THE BLACK BODY LIMIT}

One of the striking features of the BBL is the suppression of nondiagonal
transitions in the photon interaction with heavy nuclei \cite {Gribov}.
Indeed,  in the BBL  the dominant contribution to the coherent  diffraction
originates from ``a shadow'' of the fully absorptive interactions
at impact parameters $b\leq R_A$, so  the orthogonality argument 
is applicable. We use it to derive 
the BBL expression for the differential cross section of the 
production of the invariant mass $M^2$  for scattering of (virtual)
photons~\cite{BBL}. For the real photon case:
\begin{equation}
{{d\sigma_{(\gamma +A\to ``M''+A)}}
\over dt dM^2} ={\alpha_{em} \over 3 \pi}{(2\pi R_A^2)^2\over 16\pi}
{\rho(M^2)\over M^2} {4\left|J_1(\sqrt{-t}R_A)\right|^2
\over -t R_A^2}.
\label{ccsb}
\end{equation}
Here $\rho(M^2) = 
\sigma(e^+ e^-\rightarrow hadrons)/ \sigma(e^+ e^-\rightarrow \mu^+ \mu^- )$.
Comparison of  the measured cross section of the diffractive
production of states with certain masses with the BBL result 
(Eq.~\ref{ccsb})  would allow to determine up to what masses in
the photon wave function interaction remains black.
Similar equation is valid in the BBL for the production of specific
hadronic or quark-gluonic final states ($q\bar q, q\bar q g,$ etc)
in the case of the coherent nuclear recoil. This allows
to measure  any component of the light 
cone photon wave function which interacts with the BBL
strength 
 corresponding  final states in the coherent processes. 
The onset of BBL limit for hard processes should reveal itself also in a 
faster increase with energy of cross sections of photoproduction of
excited states as compared  to the cross section  for the ground state meson.
It would be especially advantageous for these studies
to use a set of nuclei - one in the medium range, like $Ca$,  and
another with $A\sim 200$. This would allow to 
 remove the edge effects and use the
length of about 10 fm of nuclear matter.

One especially interesting channel is exclusive 
diffractive dijet production by real photons.
One may expect that for the $\gamma A$ energies which will 
be available at EIC or at LHC in UPC 
 the BBL  in the scattering off heavy
nuclei would be a good approximation for the masses $M$ in the photon
wave function  up to  few GeV. This is 
the domain which is described by perturbative QCD for
${\it x}\sim 10^{-3}$ in the case of  the  proton targets,
 and  larger $x$ 
for scattering off nuclei. The condition of large longitudinal 
distances, a small longitudinal momentum 
transfer,  will be applicable in this case up to quite large values of the
produced diffractive mass.
In the BBL the dominant channel of diffraction to  large masses is
production of two jets with the total cross section given by
Eq.(\ref{ccsb}) and with a characteristic angular distribution
$(1+ \cos^2 \theta)$, where $\theta$ is the c.m. angle \cite{BBL}. On
the contrary,  in the perturbative QCD limit the
diffractive dijet production,  except the charmed dijet production,
is strongly suppressed \cite{brodsky,diehl}. The suppression
is due to the structure of coupling of the  $q\bar q$ component of the 
real photon wave function to two gluons when calculated in the lowest order in 
$\alpha_s$. As a result,  in the real photon case hard diffraction
involving light quarks is connected to production of $q\bar q g$ and 
higher states. Distribution of diffractively produced jets over invariant mass 
provides an important  test of 
the onset of BBL limit. Really, in the DGLAP/CT regime 
differential cross section of forward diffractive dijet production
should be $\propto 1/M^8$ and be dominated by charm jet production. This 
behavior is strikingly different from the BBL limit expressions of \cite{BBL}.
Thus, the dijet photoproduction should be very sensitive to the
onset of the BBL regime. 
We want to draw attention that $q \bar q$ component
of the  photon light-cone wave function can be measured in three
 independent diffractive phenomena:
in the BBL off the proton, in BBL off a heavy nuclei, in the 
CT regime where the wave function  can be measured as a 
function of the  interquark distance~\cite{FMS}.
A competing process for photoproduction of dijets off heavy nuclei is
production of dijets in $\gamma -\gamma$ collisions where the second photon
is provided by the Coulomb field of the nucleus. 
The dijets produced in this process have positive
C-parity and hence this amplitude does not  interfere
with the amplitude of the dijet production in
the $\gamma \Pomeron$ interaction which have negative C-parity.
Our estimates indicate that this process will constitute a very small
background over the wide range of energies \cite{FSZrho}.

\section{Coherent vector meson production in UPC at LHC}

Ultraperipheral collisions(UPC) of relativistic heavy ions at RHIC 
and LHC open a promising new avenue for experimental studies 
of the photon induced coherent and incoherent interactions with nuclei
at high  energies~\cite{baur,klein99}. 
 Really, the  LHC heavy ion program \cite{felix,alice}
will allow studies of photon-proton and  photon - nucleus collisions 
at the energies exceeding by far those available now at HERA for $\gamma-p$ 
scattering.

Hence,  we can analyze an opportunity to study the phenomena discussed above
combining the theory of photo induced processes in the ultraperipheral AA 
collisions with our studies of the coherent photo(electro) production
of vector mesons. 
We can use
the standard Weizsacker-Williams approximation~\cite{ww} to calculate
 the cross section integrated over
the momentum of the nucleus which emits the quasireal photons.

The  cross section of the vector meson production
integrated over the transverse momenta of the nucleus which emitted a
photon  can be written in the convoluted form:
\begin{equation}
{d \sigma(AA\to V AA)\over dy}=
{N_{\gamma}(y)}  \sigma_{\gamma A\rightarrow V A}(y)+
{N_{\gamma}(-y)}  \sigma_{\gamma A\rightarrow V A}(-y).
\label{base}
\end{equation}
Here y is the rapidity
\begin{equation}
y={1\over 2}\ln{E_V-p_3^V \over E_V +p_3^V}.
\end{equation}
The flux of the equivalent photons ${N_{\gamma}(y)}$ is given by 
a simple expression~\cite{baur}:
\begin{equation}
N(y))=\frac {Z^2\alpha} {{\pi}^2}\int d^2b \Gamma_{AA}({\vec b}) \frac{1} {b^2}X^2
\bigl [K^2_1(X)+\frac {1} {\gamma} K^2_0(X)\bigr ].
\end{equation}
Here $K_0(X)$ and $K_1(X)$ are modified Bessel functions with
argument $X=\frac {bm_Ve^y} {2\gamma}$, $\gamma$ is Lorentz factor and 
${\vec b}$ is the impact parameter. 
The Glauber profile factor 
\begin{equation}
\Gamma_{AA}({\vec b})=
exp\biggl (-\sigma_{NN}
\int \limits^{\infty}_{-\infty}dz\int d^2b_1
\rho(z,{\vec b_1})\rho(z,{\vec b}-{\vec b_1})\biggr ),
\end{equation} 
accounts for the inelastic strong interactions of
the nuclei at impact parameters  $b \le 2R_A$ and, hence, suppresses
the corresponding contribution of the vector meson photoproduction.

 Recently the STAR collaboration released the first data on 
the cross section of the coherent $\rho $-meson production in gold-gold 
UPC at $W_{NN}=\surd s_{NN}=130$ GeV~\cite{star}. This provides a first opportunity 
to check the basic features of the theoretical models and main 
approximations  which include the Weizs\"{a}cker-Williams
(WW) approximation for the spectrum of the equivalent photons, an 
approximate  procedure for removing collisions at small impact parameters
where nuclei interact strongly, and the model for the vector meson production
in the $\gamma A $ interactions. In the case of  the $\rho$-meson production
the basic process is understood much better than for
other photoproduction processes. Hence, checking the theory
for this case is especially important for 
proving that UPC could be used for learning new information 
about photon - nucleus interactions.
 Note here that the inelastic shadowing effects which 
start  to contribute at high energies still remain a few percent  
correction at energies $\le$ 100 GeV relevant for the STAR kinematics.
For LHC energy range one should account for the blackening of
interaction with nuclei. In this case cross section of inelastic 
diffraction in hadron-nucleus collisions should tend to 0.  So major impact 
for the calculation of the process of diffractive photoproduction of 
$\rho$ meson would be necessity to neglect by the contribution of 
$\rho'$~\cite{FSZrho}. 

\begin{figure}
    \centering
        \leavevmode
        \epsfxsize=0.8\hsize
        \epsffile{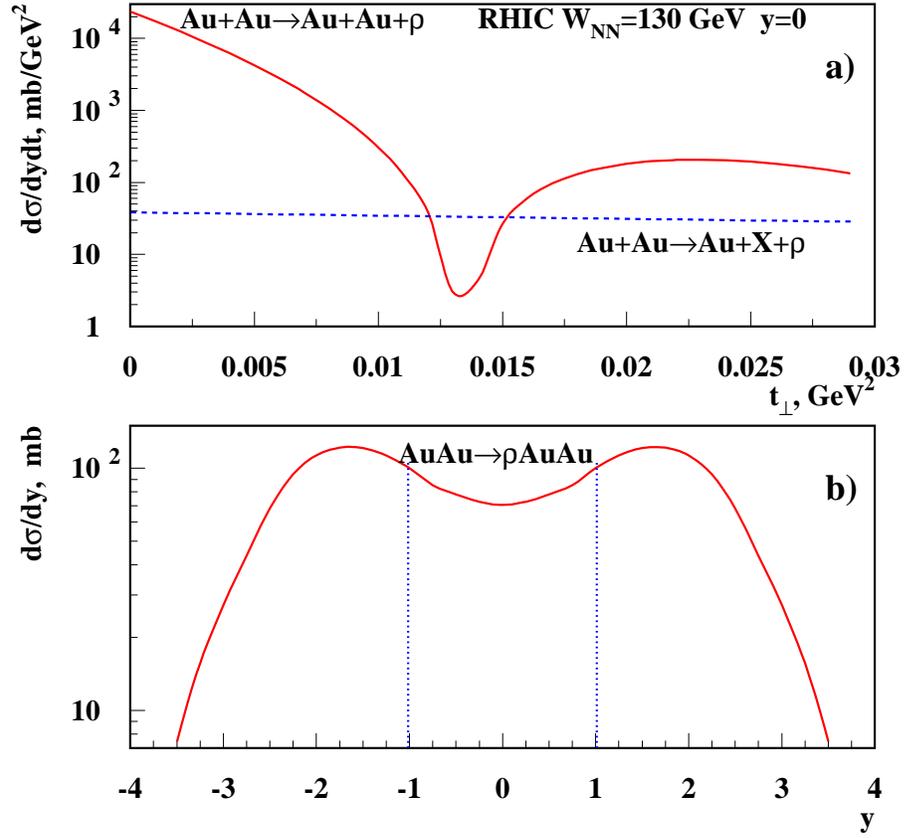}
\caption{(a) Momentum transfer dependence of the coherent and
incoherent $\rho$-meson
production in Au-Au UPC at $\surd s_{NN}=130$ GeV calculated in GVDM.
(b) Rapidity distributions for coherent
 $\rho $-meson production in the gold-gold UPC at $\surd s_{NN}=130$ GeV 
calculated in GVDM.}
\label{figstar1}
\end{figure}

The  calculated momentum transfer distributions at the rapidity $y=0$ and 
the momentum transfer integrated rapidity distribution for gold-gold UPC 
at $\surd s_{NN}=130$ GeV  are presented in 
Figs.~\ref{figstar1}a,b~\cite{fszstar}. 
\begin{figure}
    \centering
        \leavevmode
        \epsfxsize=0.8\hsize
        \epsffile{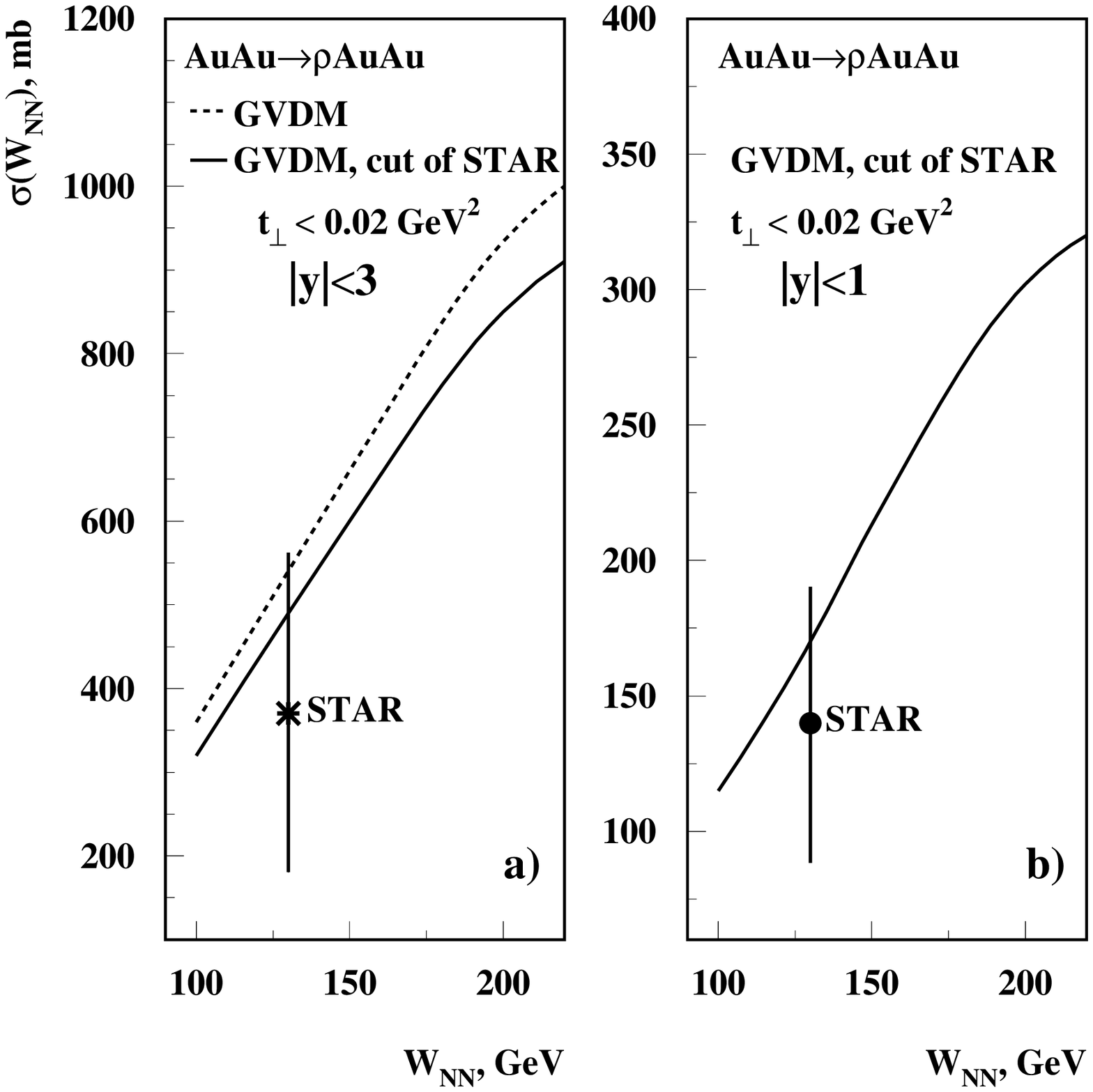}
\caption{Energy dependence of the total  cross section for coherent
 $\rho $-meson production in the gold-gold UPC calculated in the GVDM. }
\label{figstar2}
\end{figure}
Let us briefly comment on our estimate of the incoherent $\rho $-meson
production cross section. The momentum transfer distribution (dashed line 
in Fig.~\ref{figstar1}a) 
is practically flat in the discussed $t_{\bot }$ range.
The total incoherent cross section obtained by integration over 
the wide range of $t_{\bot }$ is $\sigma _{inc}=120$ mb. To select 
the coherent production the cut $t_{\bot }\leq 0.02$ GeV$^2$ was used  
in the data analysis \cite{star}. 
Correspondingly, the calculated incoherent cross  section for
this region of $t_{\bot }$ is $\sigma _{inc}=14$ mb. Our calculations 
of incoherent production which are based 
on accounting for only the single elementary diffractive collision  
obviously present the lower limit. The residual nucleus will be 
weakly excited and can evaporate only one-two neutrons.  
The events $A+A\to \rho+xn+A_{1}+A_{2}$  were detected 
by the STAR and identified as a 
two-stage process - coherent $\rho $-production
with the subsequent electromagnetic excitation and neutron 
decay of the  colliding nuclei \cite{Baltz}. In particular, the 
cross section estimated by the STAR for the  case when only one of the nuclei 
is excited and emits several  neutrons is 
$\sigma ^{\rho }_{xn,0n}=95\pm 60\pm 25$ mb.   
The momentum transfer distribution for these events is determined by
the  coherent production.  Hence, it differs from that
for incoherent events but in the region of very low  $t_{\bot }$
it is hardly possible  to separate 
them experimentally,  and  the measured cross section 
$\sigma ^{\rho }_{xn,0n}$ includes contribution
of incoherent events on the level of $15\%$.     

The total rapidity-integrated cross section of coherent $\rho $-meson
production calculated in the GVDM for the range of energies available at RHIC
is shown in Fig.~\ref{figstar2}(solid line).
We find $\sigma _{coh}^{th}=540$ mb at $\surd s_{NN}=130$ GeV. 
The  value $\sigma _{coh}^{exp}=370\pm 170\pm 80$ mb
was obtained at this energy by the STAR from 
the data analysis at the low momentum transfer $t_{\bot }\leq 0.02$ GeV$^2$. 
Thus, before making a comparison we should take into account
this cut.
It leads to a reduction of the  cross section  
by $\approx 10\%$ (the dashed line in Fig.~\ref{figstar2}).
In our calculations we didn't account for the $t_{\bot }$-dependence
of the elementary amplitudes which are rather flat in the considered
range of energies and momentum transfers 
as compared to that for the nucleus form factor.
So, in the region of integration important for our analysis
it is reasonable to neglect this slope.
Nevertheless, an account of this effect would slightly 
reduce our estimate of the total cross section. Also we neglected a smearing 
due to the transverse momentum of photons and 
the interference of the production amplitudes from both nuclei~\cite{Klein992}. 

\begin{table}[t]
\centering
\begin{tabular}{|c|c|c|}\hline
 Approximation  & Ca-Ca at LHC($\gamma =3500$)  &Pb-Pb  at LHC($\gamma =2700$) \\ \hline\hline
Impulse  &     0.6 $mb$         &   70 $ mb$          \\ \hline
Leading twist    &  0.2 $mb$  &   15 $mb$          \\ \hline
\end{tabular}
\caption{Total cross sections of $J/\psi$ production
 in UPC at LHC.}
\label{tcrsec1}
\end{table}

\begin{table}[b]
\centering
\begin{tabular}{|c|c|c|}\hline
 Approximation  & Ca-Ca at LHC($\gamma =3500$)  &Pb-Pb  at LHC($\gamma =2700$) \\ \hline\hline
Impulse  &     1.8 $\mu b$         &   133 $\mu b$          \\ \hline
Leading twist    &  1.2 $\mu b$  &   78 $\mu b$          \\ \hline
\end{tabular}
\caption{Total cross sections of $\Upsilon$ production
 in UPC at LHC.}
\label{tcrsec2}
\end{table}

This latter phenomenon results in the narrow dip in the coherent 
$t_{\bot }$-distribution at 
$t_{\bot }\leq 5\cdot 10^{-4}$ GeV$^2$. All these effects
do not influence  noticeably  the value of the $t_{\bot }$-integrated
cross section but can be easy treated and taken into account in a more 
refined analysis. 
Thus we find $\sigma _{coh}^{th}=490$ mb to be compared to the STAR 
value $\sigma _{coh}^{exp}=370\pm 170\pm 80$ mb.
Since our calculation does not have any free parameters, this can be
considered as  a reasonable agreement.

\begin{figure}
\centering
\epsfxsize=1.0\hsize
    \epsffile{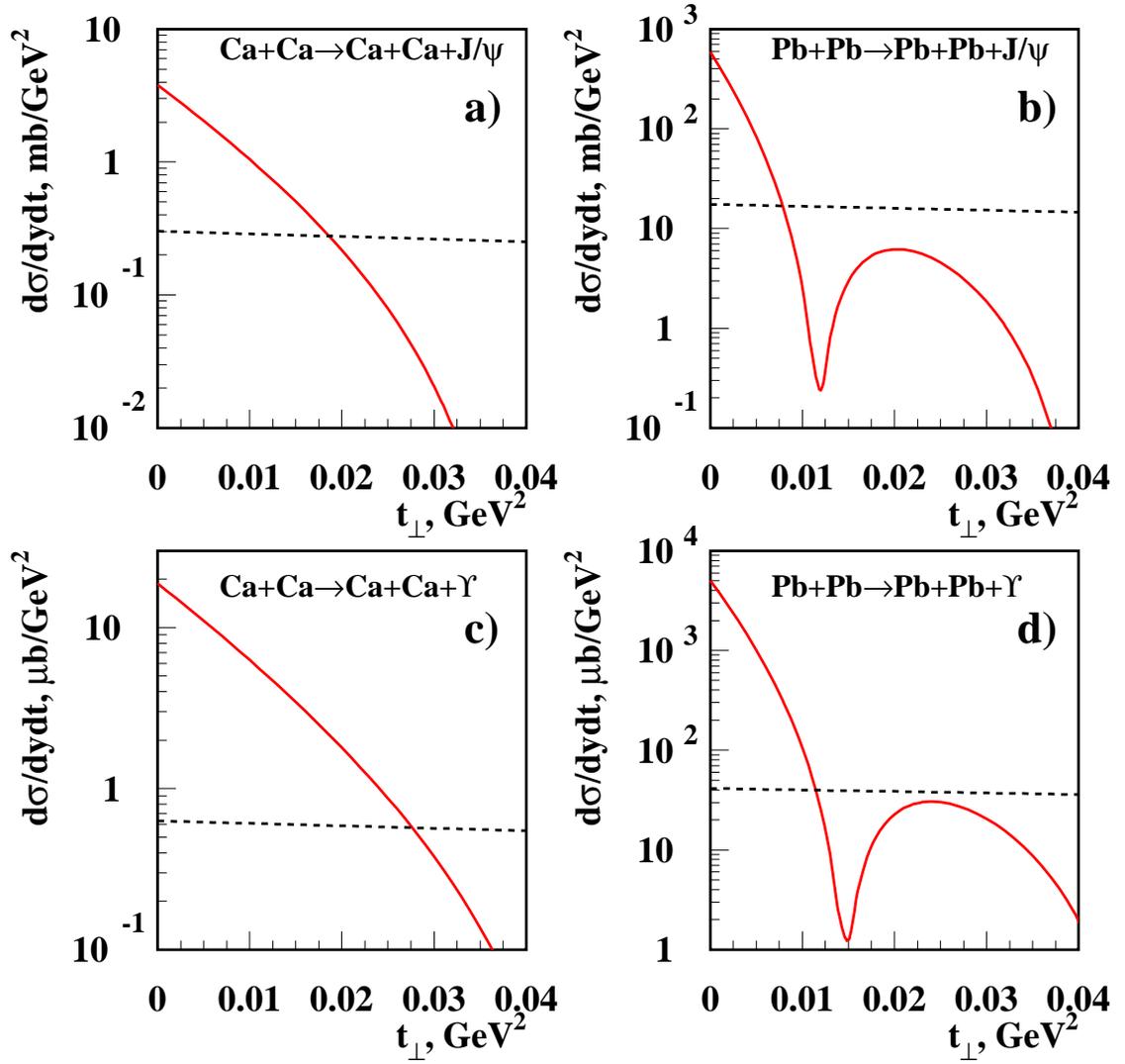}
\caption{The momentum transfer distribution for the coherent
$J/\psi$ and $\Upsilon$ production in Ca-Ca and Pb-Pb in UPC at LHC. 
Solid line - calculation with the Leading twist shadowing,
dashed line - the momentum transfer distribution for the
incoherent photoproduction.}
\label{lhcdst}
\end{figure}

\begin{figure}
\centering
        \leavevmode
        \epsfxsize=1.0\hsize
       \epsffile{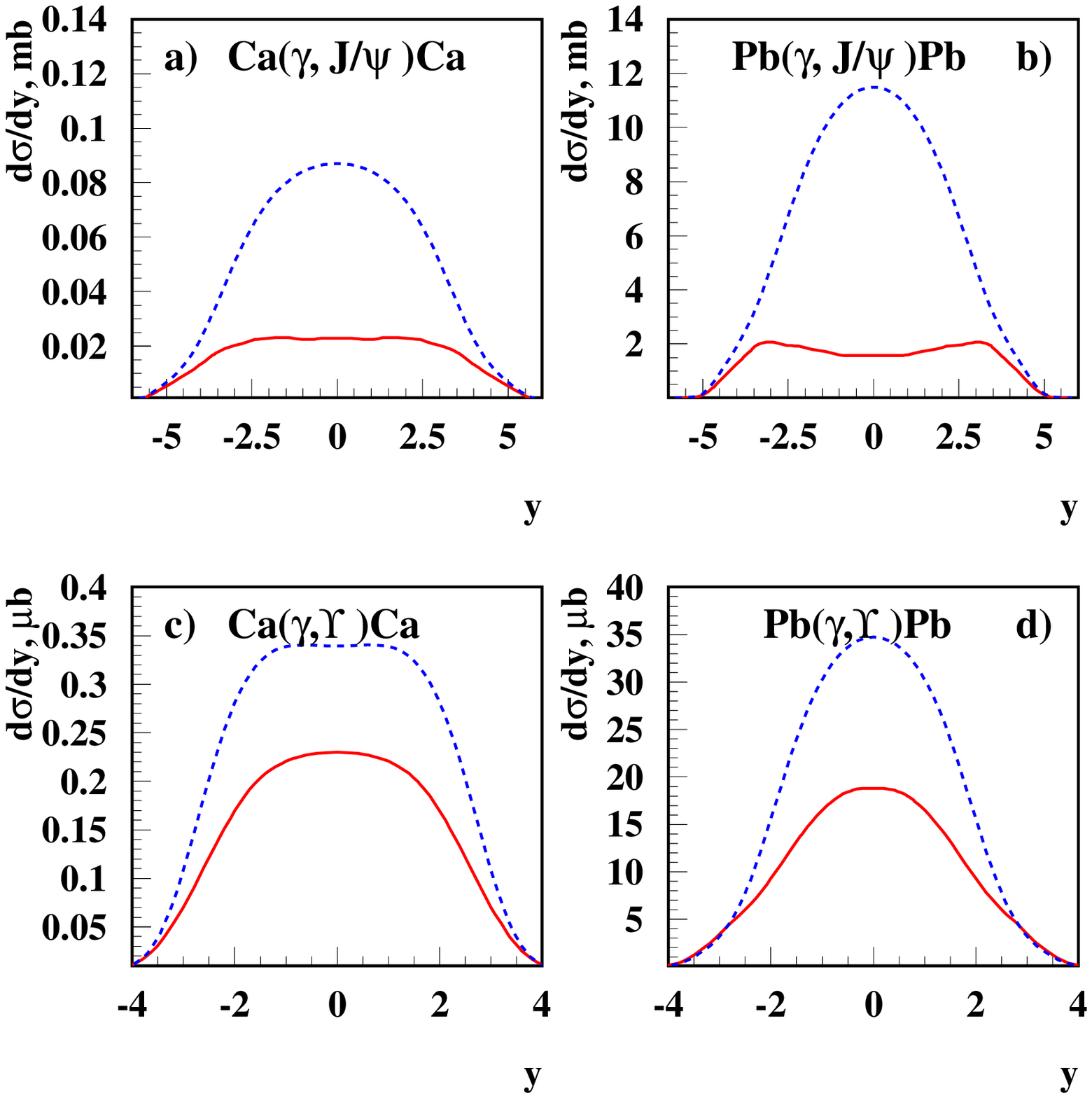}
\vspace*{1cm}
\caption{
The rapidity distributions for the $J/\psi$ and $\Upsilon$ coherent 
production off Ca and Pb in UPC at LHC calculated with  the leading 
twist shadowing based on H1 parameterization
of gluon density(solid line) and in the Impulse Approximation(dashed line)}.
\label{figlhc1}
\end{figure}

It was suggested in \cite{FKS,FS99} to look for color opacity phenomenon
using $J/\psi $ (photo) electroproduction. This however requires energies
much larger than those available at the fixed target facilities and
would require use of electron-nucleus colliders. At  the same time
estimates of the counting rates performed within the framework of the
FELIX study \cite{felix} have demonstrated that the effective photon
luminosities  generated in peripheral heavy ion collisions
at LHC would lead to significant  rates of coherent photoproduction
of vector mesons including $\Upsilon$  in reaction
\begin{equation}
 A + A \to A + A + V.
\label{react}
\end{equation}
As a result it would be possible to study at LHC 
photoproduction of vector mesons in Pb-Pb and Ca-Ca collisions
at energies much higher than the range $W_{\gamma p}\le 17.3$ GeV covered 
at the fixed target experiment at FNAL \cite{Sokoloff}. 
Note that even current  experiments at RHIC
($W_{\gamma p}\le 25$ GeV) should also  exceed this limit.
As it is clearly indicated by the STAR study the coherent photoproduction,
leaving the both interacting nuclei intact,
can be reliably identified by using the veto triggering from the two-side Zero
Degree Calorimeters which select the events not comprising the
escaped neutrons.    
 The additional requirement which enables to remove contribution of the incoherent
events with the residual nucleus in the ground state is selection of the 
produced quarkonium with small transverse momentum.
In Fig.~\ref{lhcdst} we compare
the momentum transfer distributions
for the coherent $J/\psi$ and $\Upsilon$ photoproduction 
 calculated in the Leading Twist
shadowing model with the corresponding distributions for incoherent photoproduction.
Note that we estimated the upper limit of incoherent cross section simply as
the free elementary cross section on the nucleon target multiplied by the
number of nucleons A.

In  Fig.~\ref{figlhc1}a,b we
present the rapidity distributions of the $J/\psi$ coherent production
for peripheral collisions at LHC calculated  including effects of
gluon shadowing and in the impulse approximation. 
At the central rapidities we find suppression by a factor 4 for a case of Ca
and more strong, by a factor 6 for Pb.
The total cross 
sections are given in Table~\ref{tcrsec1}. 
 The rapidity distributions for coherent $\Upsilon$ production in the
UPC with Ca and Pb beams are shown in Fig.~\ref{figlhc1}c,d 
and the corresponding total cross sections
are given in Table \ref{tcrsec2}.
As it is seen from comparison of the Leading Twist shadowing based calculations to that
performed in the Impulse Approximation 
the yield of $\Upsilon$ is expected to be suppressed by a factor 2 
at central rapidities due
to the leading twist nuclear shadowing.
 
Hence, study of the coherent photoproduction of the heavy quarkonium states at LHC
opens an important avenue for investigating
the nuclear gluon distributions and the shadowing effects in the 
kinematics which would be very hard to probe
 in any other experiments.

\section{CONCLUSIONS}
We demonstrated that coherent diffraction off nuclei provides an
effective method of probing a possible onset of BBL regime in hard processes at 
small ${\it x}$. We predict a significant  increase of the ratio of the 
yields of $\rho,\rho^{\prime}$ mesons in coherent processes off heavy 
nuclei due to the blackening of the soft QCD interactions in which 
fluctuations of the interaction strength are present. An account of 
nondiagonal transitions leads to a prediction of a significant enhancement 
of production of heavier diffractive states especially production of high 
$p_t$ dijets. Study of these channels may allow to get an important 
information on the onset of the black body limit in the diffraction of real 
photons. We argued that  the  fluctuations of strengths of 
interactions has been observed at intermediate energies in the diffractive 
photoproduction of vector mesons. We discuss the opportunity to look for
the transition from the nuclear color transparency to the regime of the
color opacity in the ultrarelativistic peripheral ion collisions at LHC. 

We thank J.Bjorken,  A.Mueller, G.Shaw, members of the UPC study
 group,
 and our coauthors 
M.McDermott, A.Freund, V.Guzey, L.Gerland, 
 for  useful discussions and GIF and
DOE for support. M.S.  and M.Z. thank INT for hospitality
while this work was completed.


\begin{thebibliography}{99}
\bibitem{felix}FELIX Collaboration (E. Lippmaa et al.). CERN-LHCC-97-45, LHCC-I10;\\
A.~Ageev {\it et al.},
%``A Full acceptance detector at the LHC (FELIX),''
J.\ Phys.\ G {\bf 28}, R117 (2002).


\bibitem{baur2002}
G.~Baur, K.~Hencken, D.~Trautmann, S.~Sadovsky and Y.~Kharlov,
%``Coherent gamma gamma and gamma A interactions in very peripheral  collisions at relativistic ion colliders,''
Phys.\ Rept.\  {\bf 364}, 359 (2002);

[arXiv:hep-ph/0112211].

\bibitem{alice}
S.~Beole {\it et al.}  [ALICE Collaboration],
%``ALICE technical design report: Detector for high momentum PID,''
CERN-LHCC-98-19

\bibitem{FS99}L.~Frankfurt and M.~Strikman,
%``Diffraction at HERA, color opacity and nuclear shadowing,''
Eur.\ Phys.\ J.\ A {\bf 5}, 293 (1999).

\bibitem{RL}L.McLerran and R.Venugopalan, Phys.Rev. {\bf D50}, 
  225 (1994);\\                             
Phys.Rev.{\bf D59},                                                                                    
094002 (1999).   


\bibitem{Mueller}
A.~H.~Mueller,
%``Virtual Pair Creation In A Strong Bremsstrahlung Field: A QED Model
%For Parton Saturation,''
Nucl.\ Phys.\ B {\bf 307}, 34 (1988).



Y.~V.~Kovchegov,
%``Quantum structure of the non-Abelian Weizsaecker-Williams field for
%a  very large nucleus,''
Phys.\ Rev.\ D {\bf 55}, 5445 (1997);

[arXiv:hep-ph/9701229].




\bibitem{BBL}
L.~Frankfurt, V.~Guzey, M.~McDermott and M.~Strikman,
Phys.\ Rev.\ Lett.\  {\bf 87}, 192301 (2001); 

[arXiv:hep-ph/0104154]. 



\bibitem{FSZpsi}
L.~Frankfurt, M.~Strikman and M.~Zhalov, Phys.\ Lett.\ B {\bf 540}, 220 (2002);

%``Fading out of J/psi color transparency in high energy heavy ion 
% peripheral collisions,''
[arXiv:hep-ph/0111221].

\bibitem{FSZrho}
L.~Frankfurt, M.~Strikman and M.~Zhalov,
Phys.\ Lett.\ B {\bf 537}, 51 (2002).
%``Signals for black body limit in coherent ultraperipheral 
%heavy ion  collisions,''


\bibitem{fszstar}
L.~Frankfurt, M.~Strikman and M.~Zhalov,  Phys.\ Rev.\ {\bf C67}, 034901 (2003);

%``Predictions of the generalized Glauber model for the coherent  rho
%production at RHIC and the STAR data,''
[arXiv:hep-ph/0210303].
%%CITATION = HEP-PH 0210303;%%


\bibitem{fszferrara}
L.~Frankfurt, M.~Strikman and M.~Zhalov,
Nucl.\ Phys.\ {\bf A711}, 243 (2002).

\bibitem{fgszch}
L.~Frankfurt, L.~Gerland, M.~Strikman and M.~Zhalov,
%``Cross section oscillations in the coherent charmonium photoproduction  
%off nuclei at moderate energies,''
[arXiv:hep-ph/0301028].


\bibitem{fgszloch}
L.~Frankfurt, L.~Gerland, M.~Strikman and M.~Zhalov,
%``Probing coherent charmonium photoproduction off light nuclei at medium  energies,''
[arXiv:hep-ph/0301077].

\bibitem{fgszups}L.~Frankfurt, V.~Guzey, M.~Strikman and M.~Zhalov, 

[arXiv:hep-ph/0304218]

\bibitem{glauber}
R.~J.~Glauber, Boulder Lectures in Theoretical Physics, vol.1, \\
(Interscience Publ.\ Inc.\ , NY)(1959);\\
K.~Gottfried and D.~R.~Yennie,
%%``Vector Mesons And Nuclear Optics,''
Phys.\ Rev.\  {\bf 182}, 1595 (1969)\\
Erratum-ibid.\ D {\bf 2}, 2737 (1970);\\ 
~G.~V.~Bochmann, Phys.~Rev.~ D6, 1938 (1972).


\bibitem{yenn}
T.~H.~Bauer, R.~D.~Spital, D.~R.~Yennie and F.~M.~Pipkin,
%``The Hadronic Properties Of The Photon In High-Energy Interactions,''
Rev.\ Mod.\ Phys.\  {\bf 50}, 261 (1978);

[Erratum-ibid.\  {\bf 51}, 407 (1978)].


\bibitem{Gribov} V. N.Gribov,  Zh. Eksp. Teor. Fiz.
 \textbf{57}, 1306 (1969).


\bibitem{Brodsky}
S.~J.~Brodsky and J.~Pumplin,
%``Photon - Nucleus Total Cross-Sections,''
Phys.\ Rev.\  {\bf 182}, 1794 (1969).


\bibitem{GVDM}A. Donnachie and G. Shaw, in
Electromagnetic Interactions of Hadrons, edited by \\A.
 Donnachie and G. Shaw
    (Plenum, New York, 1978), Vol. 2, pp. 164-194;\\
 H. Fraas, B. Read, and D. Schildknecht,
Nucl.\ Phys.\ {\bf B86} , 346 (1975);\\
P.~Ditsas and G.~Shaw,
%``Shadowing In Nuclear Photoabsorption,''
Nucl.\ Phys.\ B {\bf 113}, 246 (1976).

\bibitem{Gribovinel}V.~N.~Gribov,
Sov.\ J.\ Nucl.\ Phys.\  {\bf 9}, 369 (1969);\\
Sov.\ Phys.\ JETP {\bf 29},483 (1969);\\
Sov.\ Phys.\ JETP {\bf 30}, 709 (1970).

\bibitem{FS88}L. L. Frankfurt and M. Strikman, Phys. Rep. {\bf 160}, 235 (1988).

\bibitem{FGS97}
L.~Frankfurt, V.~Guzey and M.~Strikman,
%``Cross section fluctuations of photon projectile in
% generalized vector  meson dominance model,''
Phys.\ Rev.\ D {\bf 58}, 094039 (1998).

\bibitem{HFS}
M.~Beiner, H.~Flocard, N.~Van Giai, P.~Quentin,
Nucl.\ Phys. {\bf A238}, 29 (1975).

\bibitem{p2p}
S.~L.~Belostotsky { et al.},
%``Quasielastic Proton Scattering At 1-Gev,''
 Proceedings of the conference
Modern developments in nuclear physics,\\ Novosibirsk 1987, World Scientific, 
1988, p. 191.

\bibitem{eep}
L.~Lapikas, G.~van der Steenhoven, L.~Frankfurt,
M.~Strikman and M.~Zhalov,\\
%``The transparency of C-12 for protons,''
Phys.\ Rev.\ C {\bf 61}, 064325 (2000);

[arXiv:nucl-ex/9905009];

L.~Frankfurt, M.~Strikman and M.~Zhalov,
%``Single particle strength restoration and
%nuclear transparency in high  Q**2 exclusive (e,e p) reactions,''
Phys.\ Lett.\ B {\bf 503}, 73 (2001);

[arXiv:hep-ph/0011088].


\bibitem{Pautz:qm}
A.~Pautz and G.~Shaw,
%``Nuclear Shadowing And Rho Photoproduction,''
Phys.\ Rev.\ C {\bf 57}, 2648 (1998).


\bibitem{MIT}
H.~Alvensleben et al.,
Nucl.\ Phys.\ {\bf B18}, 333 (1970);\\
G.~McClellan et al.,
Phys.\ Rev.\ {\bf D4}, 2683 (1971).


\bibitem{LD}
A.~Donnachie and P.~V.~Landshoff,
%``Exclusive vector photoproduction: Confirmation of Regge theory,''
Phys.\ Lett.\ B {\bf 478}, 146 (2000);

[arXiv:hep-ph/9912312].

\bibitem{GW}M.~Good and W.~Walker, Phys. Rev. {\bf D 120}, 1857 (1960).


\bibitem{Badelek:gs}
B.~Badelek, M.~Krawczyk, K.~Charchula and J.~Kwiecinski,
%``Small X Physics In Deep Inelastic Lepton Hadron Scattering,''
Rev.\ Mod.\ Phys.\  {\bf 64} (1992) 927.
\bibitem{hera}
H.~Abramowicz and A.~Caldwell,
%``HERA collider physics,''
Rev.\ Mod.\ Phys.\  {\bf 71}, 1275 (1999);

[arXiv:hep-ex/9903037].
%%CITATION = HEP-EX 9903037;%%

\bibitem{FGMS2002}
L.~Frankfurt, V.~Guzey, M.~McDermott and M.~Strikman,
%``Nuclear shadowing in deep inelastic scattering on nuclei: Leading twist
%versus eikonal approaches,''
JHEP {\bf 0202}, 027 (2002);

[arXiv:hep-ph/0201230].
%%CITATION = HEP-PH 0201230;%%


\bibitem{slac2}
U.~Camerini et al., Phys.\ Rev.\ Lett.\  {\bf 35}, 483 (1975).

\bibitem{FS91}
L.~Frankfurt and M.~Strikman,
%``Color Screening And Color Transparency In Hard Nuclear Processes,''
Prog.\ Part.\ Nucl.\ Phys.\  {\bf 27}, 135 (1991).
%%CITATION = PPNPD,27,135;%%
\bibitem{kophuf}
J.~Hufner, B.~Kopeliovich, Phys. Lett. {\bf B426}, 154 (1988).


\bibitem{FKS}
L.~Frankfurt, W.~Koepf and M.~Strikman,
%``Diffractive heavy quarkonium photo- and electroproduction in QCD,''
Phys.\ Rev.\ D {\bf 54}, 3194 (1996),ibid {\bf 57}, 512 (1998).

\bibitem{Suzuki:2000az}
K.~Suzuki, A.~Hayashigaki, K.~Itakura, J.~Alam and T.~Hatsuda,
%``Validity of the color dipole approximation for diffractive
%production  of heavy quarkonium,''
Phys.\ Rev.\ D {\bf 62}, 031501 (2000);

[arXiv:hep-ph/0005250].




\bibitem{vmvogt}
R.~Vogt, Phys.\ Rept.\ {\bf 310}, 197 (1999);\\
C.~Gerschel, J.~Hufner, Annu.\ Rev.\ Nuc.\ Part.\ Sci. {\bf 49}, 255 (1999).

\bibitem{kharzeev}
D.~Kharzeev and H.~Satz,
%``Color deconfinement and quarkonium dissociation,''
In R.C. HWA (ed.): Quark-gluon plasma, vol.2, 395-453 and

[arXiv:hep-ph/9505345].


\bibitem{slac1}
R.L.~Anderson et al., Phys.\ Rev.\ Lett.\  {\bf 38}, 263 (1977).


\bibitem{e160}
V.Ghazikhanian et. al. SLAC-Proposal E-160, 2000.


\bibitem{Eichten}
E.~Eichten, K.~Gottfried, T.~Kinoshita, J.~B.~Kogut, K.~D.~Lane and 
T.~M.~Yan,\\
%``The Spectrum Of Charmonium,''
Phys.\ Rev.\ Lett.\  {\bf 34}, 369 (1975);
[Erratum-ibid.\  {\bf 36}, 1276 (1976)].
%%CITATION = PRLTA,34,369;%%

\bibitem{Richard}
J.~M.~Richard,
%``Ground State Admixture Into The Psi-Prime-Prime (3.772),''
Z.\ Phys.\ C {\bf 4} (1980), 211 (1980).
%%CITATION = ZEPYA,C4,211;%%

\bibitem{fgszpsip}
L.~Frankfurt, L.~Gerland, M.~Strikman and M.~Zhalov,
%``The psi''/psi' ratio as unambiguous signature for hard physics in  
%nuclear reactions and in decays of beauty hadrons,''
[arXiv:hep-ph/0302009].




\bibitem{CFS}
J.~C.~Collins, L.~Frankfurt and M.~Strikman,
%``Factorization for hard exclusive electroproduction of mesons in QCD,''
Phys.\ Rev.\ D {\bf 56}, 2982 (1997);

[arXiv:hep-ph/9611433].



\bibitem{BFGMS}
S.~J.~Brodsky, L.~Frankfurt, J.~F.~Gunion, A.~H.~Mueller and M.~Strikman,\\
%``Diffractive leptoproduction of vector mesons in QCD,''
Phys.\ Rev.\ D {\bf 50}, 3134 (1994); 
[arXiv:hep-ph/9402283].




\bibitem{FS89}
L.~Frankfurt and M.~Strikman,
%``Large T, Triple Pomeron Diffractive Processes In QCD,''
Phys.\ Rev.\ Lett.\  {\bf 63}, 1914 (1989);

[Erratum-ibid.\  {\bf 64}, 815 (1990)].




\bibitem{Ryskin} M.~G.~Ryskin,
%``Diffractive J / psi electroproduction in LLA QCD,''
Z.\ Phys.\ C {\bf 57}, 89 (1993).




\bibitem{FMS}L.~Frankfurt, G.~A.~Miller and M.~Strikman, 
Phys.\ Lett.\ B {\bf 304}, 1 (1993).

\bibitem{Low}
F.~E.~Low,
%``A Model Of The Bare Pomeron,''
Phys.\ Rev.\ D {\bf 12}, 163 (1975).
%%CITATION = PHRVA,D12,163;%%
 

\bibitem{E791}E.~M.~Aitala {\it et al.}  [E791 Collaboration],
Phys.\ Rev.\ Lett.\  {\bf 86}, 4773 (2001).

\bibitem{Sokoloff}
M.~D.~Sokoloff {\it et al.},
Phys.\ Rev.\ Lett.\  {\bf 57}, 3003 (1986).




\bibitem{FMS2000}
L.~Frankfurt, M.~McDermott and M.~Strikman,JHEP,{\bf 0103}, 045 (2001);

[arXiv:hep-ph/0009086].
%%%psi paper

\bibitem{Rad}
A.~V.~Radyushkin, In ``At the Frontier of Particle Physics / Handbook
of QCD'', edited by M. Shifman (World Scientific, Singapore, 2001),
vol. 2*
p. 1037-1099;

[arXiv:hep-ph/0101225].

\bibitem{MF}A.~Freund and M.~McDermott, Phys.~Rev.~D {\bf 65}, 074008 (2002).

\bibitem{fmsupsilon}
L.~L.~Frankfurt, M.~F.~McDermott and M.~Strikman,
%``Diffractive photoproduction of Upsilon at HERA,''
JHEP {\bf 9902}, 002 (1999);

[arXiv:hep-ph/9812316].
%%CITATION = HEP-PH 9812316;%%

\bibitem{Martin:1999rn}
A.~D.~Martin, M.~G.~Ryskin and T.~Teubner,
%``Upsilon photoproduction at HERA compared to estimates of
%perturbative  {QCD},''
Phys.\ Lett.\ B {\bf 454}, 339 (1999);

[arXiv:hep-ph/9901420].


\bibitem{fgsrev}
L.~Frankfurt, V.~Guzey and M.~Strikman,
%``Color coherent phenomena on nuclei and the QCD evolution equation,''
J.\ Phys.\ G {\bf 27}, R23 (2001);

[arXiv:hep-ph/0010248].
%%CITATION = HEP-PH 0010248;%%



\bibitem{Collins}J.~C.~Collins,
Phys.\ Rev.\ D {\bf 57}, 3051 (1998);
[Erratum-ibid.\ D {\bf 61}, 019902 (1998)];

[arXiv:hep-ph/9709499].


\bibitem{H1}C.~Adloff {\it et al.}  [H1 Collaboration],
Eur.\ Phys.\ J.\ C {\bf 20}, 29 (2001).


\bibitem{guide} L. Frankfurt, V. Guzey and M. Strikman, preprint hep-ph/0303022.

\bibitem{Frankfurt:2002ka}
L.~Frankfurt and M.~Strikman,
%``Two-gluon form factor of the nucleon and J/psi photoproduction,''
Phys.\ Rev.\ D {\bf 66}, 031502 (2002);

[arXiv:hep-ph/0205223].

\bibitem{FP} E.~L.~Feinberg and I.~Ia.~Pomeranchuk, Suppl.\ Nuovo \
  Cimento \ {\bf III}, 562 (1956). 

\bibitem{FMSdiff}
L.~Frankfurt, G.~A.~Miller and M.~Strikman,
%``Evidence for color fluctuations in hadrons from coherent nuclear diffraction,''
Phys.\ Rev.\ Lett.\  {\bf 71}, 2859 (1993);

[arXiv:hep-ph/9309285].


\bibitem{Kaidalov}
A.~B.~Kaidalov, V.~A.~Khoze, A.~D.~Martin and M.~G.~Ryskin,
%``Diffraction of protons and nuclei at high energies,''
[arXiv:hep-ph/0303111].


\bibitem{KleinVogt}
S.~R.~Klein, J.~Nystrand and R.~Vogt,
%``Heavy quark photoproduction in ultra-peripheral heavy ion collisions,''
Phys.\ Rev.\ C {\bf 66}, 044906 (2002);

[arXiv:hep-ph/0206220].




\bibitem{AGK}V.A.  Abramovski$\breve{{\rm i}}$,                                                          
V.N. Gribov and O.V. Kancheli, Yad. Fiz. {\bf 18}, 595 (1973).

\bibitem{FSAGK} L.Frankfurt and M.Strikman, Phys.Lett. {\bf B382}, 6 (1996).


\bibitem{Dino}
K.~Goulianos  [CDF Collaboration],
%``Multigap diffraction at CDF,''
Acta Phys.\ Polon.\ B {\bf 33} (2002) 3467
[arXiv:hep-ph/0205217].

\bibitem{STZ}
M.~Strikman, M.~G.~Tverskoy and M.~B.~Zhalov,
%``Soft neutron production in DIS: A window to the final state  interactions,''
Phys.\ Lett.\  {\bf B459}, 37 (1999);

[arXiv:nucl-th/9806099].



\bibitem{brodsky}
S.~J.~Brodsky and J.~Gillespie,
%``Second Born Corrections To Wide Angle Electron
%Pair Production And Bremsstrahlung,''
Phys.\ Rev.\  {\bf 173}, 1011 (1968).



\bibitem{diehl}
M.~Diehl,
%``Diffractive production of dijets at HERA,''
Z.\ Phys.\ C {\bf 66}, 181 (1995).





\bibitem{baur}
G.~Baur, K.~Hencken, D.~Trautmann, S.~Typel and H.~H.~Wolter,
%``Electromagnetic Dissociation as a Tool for Nuclear Structure and Astrophysics,''
Prog.\ Part.\ Nucl.\ Phys.\  {\bf 46}, 99 (2001);

[arXiv:nucl-th/0011061].

\bibitem{klein99}S.~Klein, J.~Nystrand,
Phys.\ Rev.\ C {\bf 60}, 014903 (1999);

[arXiv:hep-ph/9902259].



\bibitem{ww}
E.~Fermi, Z.\ Physik, {\bf 29}, 315 (1924);\\
C.~F.~von~Weizsacker, Z.\ Physik,\ {\bf 88}, 612 (1934);\\ 
E.~J.~Williams\ Phys.\ Rev.,\ {\bf 45}, 729 (1934). 



\bibitem{star}
C.~Adler {\it et al.}  [STAR Collaboration],
%``Coherent rho0 production in ultra-peripheral heavy ion collisions,''
Phys.\ Rev.\ Lett.\  {\bf 89}, 272302 (2002);

[arXiv:nucl-ex/0206004].




\bibitem{Baltz}
A.~J.~Baltz, S.~R.~Klein and J.~Nystrand,
%``Coherent vector meson photoproduction with nuclear breakup in  relativistic heavy ion collisions,''
Phys.\ Rev.\ Lett.\  {\bf 89}, 012301 (2002);

[arXiv:nucl-th/0205031].

\bibitem{Klein992}
S.~R.~Klein, J.~Nystrand,
%``Interference in exclusive vector meson
%production in heavy ion  collisions,''
Phys.\ Rev.\ Lett.\  {\bf 84}, 2330 (2000).





\end{thebibliography}
\end{document}